# Main Belt Asteroid Collision Histories:
# Cratering, ejecta, erosion, catastrophic dispersions, spins, binaries, tops, and wobblers

Keith. A. Holsapple

Medina, WA, USA

98039

kholsapple@comcast.net

Key Words: Asteroids; Asteroids, rotation; Collisional physics; Impact processes; Rotational dynamics

Manuscript has 66 pages, including 29 Figures and 2 Tables.

# ABSTRACT

This is a study of the collisional history of the asteroids in the main asteroid belt. Over the ±4.5 Byr of its existence, every asteroid has been impacted by others a multitude of times, producing cratering, erosion, spin, fragmentation, and occasional catastrophic disruption and dispersion. Extensive information for asteroid orbits, sizes, shapes, composition, and rotation states of those asteroids is now available. Those are a result of their history, but to interpret them requires understanding the processes. That understanding can be improved by simulations of the history.

A simulation needs robust models of the dynamical and collisional events. Such models have evolved substantially in the last few decades. Here I present current models, a method, and a code *"SSAH"* (Stochastic Simulations of Asteroid Histories) for statistical recreations of the collisional history of the main belt. Although there are still significant holes in our understanding of the appropriate models, the code exposes those and gives a framework upon which existing and future models can be tested.

The results reveal new paradigms for asteroid histories, including the distribution of spins; the irrelevance of material strength spin limits; the 'uncommon' spins of such asteroids as the rapid rotating 2001 OE84, and of large slow spinning, tumbling object Mathilde; the "V-shape" in the spin versus diameter plot; the non-Maxwellian distribution of spins of a given diameter range; the numbers of expected binaries and of tumblers, and more. They suggest a reassessment of the importance of the role of the YORP processes. At the same time, the simulations expose gaps in our knowledge that require further research. The intent here is to choose plausible models, simulate the histories, discuss the uncertainties, and provide a framework for future improvements.

The *SSAH* code is freely available for the use of others.

# 1. INTRODUCTION.

Asteroids in the main belt are subjected to processes which determine the history of the states of those asteroids. Those processes include multiple collisions producing cratering, catastrophic dispersions, pulverizations, fragmentation, spin changes, gravitational and Yarkovsky driven trajectory evolution, YORP, and perhaps others.

Dohnanyi, 1969 studied the evolution of the population-size distribution of the belt; especially on conditions for which that distribution is invariant in time. Dohnanyi's analysis of the population evolution was based on a differential equation for the time rate of the number of elements in each bin of a binned population description. That equation has three terms: 1) those objects lost from a bin due to catastrophic disruptions, 2) the number of objects gained in a bin due to fragmentation of larger objects and 3) bodies lost via dynamical processes. Each one of those terms was estimated from an expected average in each bin of the population diameters.

Davis, 1979, and Davis et al., 1985 advanced that approach with a new collisional evolution code, adding improvements for the impact disruption scaling law. In the subsequent several decades, several groups (Davis et al., 1989, Durda and Dermott, 1993, Campo Bagatin et al., 1994, Campo Bagatin et al., 2001, Durda and Dermott, 1997, Durda et al., 1998, Marzari et al., 1995, Marzari, 1999, O'Brien and Greenberg, 2005, Cheng, 2004, W. F. Bottke et al., 2005, W. Bottke et al., 2005) also investigated the collisional evolution of the main belt. They remained focused on the evolution of the population and used binned models and differential equations for the binned population, but with improved impact models. Others used similar methods to investigate small body populations in the outer Solar System, e.g. Stern, 1996, and Charnoz and Morbidelli, 2003. A primary motivation of all these studies was to explore the effects of different asteroid collisional disruption ('CD', or $Q^*$) laws, different starting main belt populations, and, in some cases, different fragmentation laws (e.g., Petit and Farinella, 1993; Campo Bagatin et al., 2001).

Those studies did not consider the spins of the asteroids. Harris, 1979 and Dobrovolskis and Burns, 1984 studied the evolution of the spins of asteroids due to small cratering impacts. Cellino et al., 1990 considered the larger impacts comparable to the CD threshold. Farinella et al., 1992 incorporated that theory into the population evolution model of Davis et al. 1989. They made the important observation that asteroid evolution is dominated by large near-catastrophic events and not by the multitude of small cratering events. However, their spin results did not match the known data.

Two decades ago Paolicchi et al., 2002 stated: "There is no satisfactory model of the collisional evolution of asteroids that includes spin properties"; and "The creation of a robust theoretical scenario for the evolution of rotational properties of asteroids is a rather ambitious task facing future researchers." More recently, Chang et al., 2019 stated: "a comprehensive simulation on the spin-rate evolution for the entire main asteroid belt is needed …".

Such a simulation is the topic of this paper.

Over the last two decades, two additional processes regarding asteroid histories have been identified and popularized: the thermal processes of Yarkovsky caused by sunlight which can change asteroid orbits, and of YORP which affects the spin of irregular shaped objects.

So, at present, a complete model for the history of an asteroid should include orbital dynamics; Yarkovsky; YORP; and the collisional physics with cratering, impact dispersions, pulverizations, and fragmentation; dynamic depletion, and perhaps other processes. Then, after some assumptions about an initial state at some time (perhaps after an early accretion phase), one could track the evolution of the asteroids and solve for the current states and population size-frequency distribution (SFD) to compare to the observations.

There are interactions between the various processes. The impact mechanics are probably more-or-less independent of spin states, which makes them independent of YORP effects. However, the reverse is certainly not true: YORP depends on surface morphology and is likely changed by every impact, although we do not have the information nor models to construct deterministic effects.

The main belt asteroid population is affected by the fragmentation and dynamic depletion models, which affect the subsequent collisional histories. But it is common to think that the population has been more-or-less steady over the last several billion years. Indeed, a primary goal of the Dohnanyi-type studies is to discover combinations of the three-part models of the Dohnanyi type that result in a steady population. That requires a balance between impact effects, dynamic dispersion and fragmentation models.

However, when the result is a steady population, we do not need to know why. We can jump to the end and simply <u>assume</u> that the population is steady[1] and simulate the collisional histories. In that case the tracking of the collisional history becomes much simpler. That is the approach of this study. I adopt the current population SFD according to observations and assume it has been steady for some time. Using that, I simulate the collisional evolution of the asteroids.

The simulations here use a stochastic model of the histories of a set of individual asteroids. That 'target' set is chosen by the user. Their impactors are defined separately by statistically drawing from a steady main belt population. The analysis exploits a time-stepping simulation of the selected set of asteroid targets with random properties and is executed in a code named *'SSAH*". *SSAH* allows an impactor population that varies in time, although only the steady population case is used here. As objects are struck, their sizes, shapes, and spin states change. The code allows output to be recorded at each time step for later 'movie' outputs. It is written in the '*Wolfram'* code language of the '*Mathematica'* program, which is required to run it.

In each time step each sample target will be impacted by many thousands or even millions of impactors of various sizes: from objects almost the size of the target, down to centimeter sizes. To make the calculation practical, an explicit-implicit scheme is used. Each impact considered "significant" (the larger ones) is modeled explicitly, one by one, using the current models of impact mechanics. The remainder of the impacts, the smallest but least important ones, are considered implicitly using binned-average effects like the historical studies mentioned above. That scheme allows one to track all the explicit effects of the largest impacts; those would be lost in studies that only calculated bin averages.

---

[1] In fact, the assumption of a steady population over a long time is mostly important for large objects (> 100 km diameter) which survive the entire epoch of the asteroid belt history. The small objects do not exist long compared to the Solar System history, and their present states are determined only by the very recent history of the population. For that reason, the restriction to a steady population for time intervals much longer than their lifetimes is not important for solving for their present states for much of the size range.

The next § 2 presents properties of the main asteroid belt, including the definition of the measure of an impactor's severity, and some average statistical results about the histories of conditions in the main belt. § 3 introduces and explains the variety of impact models used in this study. § 4 presents concepts of 3D spins and wobbles. §5 describes the *SSAH* code implementations of the models. Then the variety of results are presented, the basic ones in §6 and additional applications in § 7.

The study includes objects both from the main belt and from the NEO space. That is discussed in § 8. The § 9 discusses YORP processes, which are not included in the simulations. The results for the simulations with only impact effects casts doubts on the perceived importance of those processes. The summary is written in §10. Details about the *SSAH* code are given in Appendix A.

## 2. THE MAIN ASTEROID BELT ENVIRONMENT

The central feature of the history of an asteroid in the main belt is the multitude of impacts from other asteroids, and the breakup and dispersion from severe impacts. A simulation requires a model for the impactor sizes and rates, and for the measure of each impact severity. The impactor sizes and rates are defined by the number density and velocities of the objects in the main belt; the severity measure comes from the measure of the specific energy required to catastrophically disrupt *('CD')* and disperse a given target: the so-called *Q\** curve.

### *2.1. The Size-Frequency Distribution*

The asteroid main belt contains an enormous number of objects ranging in size from the largest, 940km Ceres, down to small, perhaps sub-centimeter-sized objects. The small sizes dominate the number count. There are perhaps $10^{12}$ to $10^{13}$ with a diameter greater than 1-m but maybe $10^{18}$ with diameters larger than 1 cm.

The number- size distribution is defined by the 'Size-Frequency-Distribution' (SFD) curve: the number function $N_{>D}$ of asteroids greater than a given diameter *D* versus that diameter *D*. Many estimates of the asteroid SFD population curve have been presented in the literature over the years. Typically, over at least a portion of the diameter sizes, the cumulative and differential distributions are modeled as a power law[2]:

$$N_{>D} \sim D^{-p+1}, \ dN \sim dD^{-p}d(D). \tag{1}$$

---
[2] Colwell, 1993 and O'Brien and Greenberg, 2005 relate the variety and the relations between different forms of population power-laws.

The exponent *p* is called the differential power slope. An important reference value is $p = 3.5$, the value derived by Dohnanyi, 1969 for a steady-state population using certain (but now obsolete) impact models.

It is now generally accepted that the data of Jedicke et al., 2002 for asteroids larger than a few km is accurate and complete. Data for asteroids between about 1 km to 8 km is reported in the SKADS survey of Gladman et al., 2009 and below 1 m in Heinze et al., 2019.

Theory can offer some guidance for those smaller ones. O'Brien and Greenberg, 2005, generalizing the approach of Dohnanyi, 1969, show that in equilibrium the slope of the population curve is determined by the slope of the catastrophic disruption $Q^*$ curve for the smaller bodies in the strength regime. Regrettably, that disruption-curve slope itself is not well constrained. For the nominal case here, the nominal disruption slope is taken as -0.25. Using the results of O'Brien and Greenberg, that would give an incremental population slope of -3.6, and a cumulative population slope of -2.6. However there remains great uncertainty.

It is difficult to translate data over ranges of H-magnitude and R-band magnitudes to diameter ranges. In addition, Bottke et al., 2020 question the Gladman et al., 2009 data. Instead, they present eight different SFD curves by solving their population evolution models for a 'best fit' SFD corresponding to eight different choices of the $Q^*$ curve[3]. Those all use the Jedicke et al., 2002 data for D>1, and all have a dip centered at about *D*=0.1 km, below the size with definitive data. Being based on a model, the data for the smaller asteroids could be considered somewhat tenuous.

The first simulations here were done using the one provided by Bottke et al., 2020 as their #1population. However, it was found that the dip centered at about *D*=0.1 km consistently produced spins for asteroids around *D*=10 km that were on average too low. That feature is discussed in detail below. A second SFD was constructed with a smaller dip, and the spin results were significantly improved (§6.3.2). Specifically, that one uses the Jedicke et al., 2002 data for *D*>1, a cumulative power of -2. from 1> *D* > 0.04 and a power of -2.6 for 0.04> *D*. Finally, following the lead of Opik [1951] I assume that the Poynting-Robertson effect has removed bodies of less than 10-cm diameter. That is the SFD used here. The **Fig. 1** shows the Bottke et al. SFD and the revised one labeled KAH-SFD.

---

[3] Some have slope values for small asteroids in the strength regime that are outside physics-based theories. I would reject those.

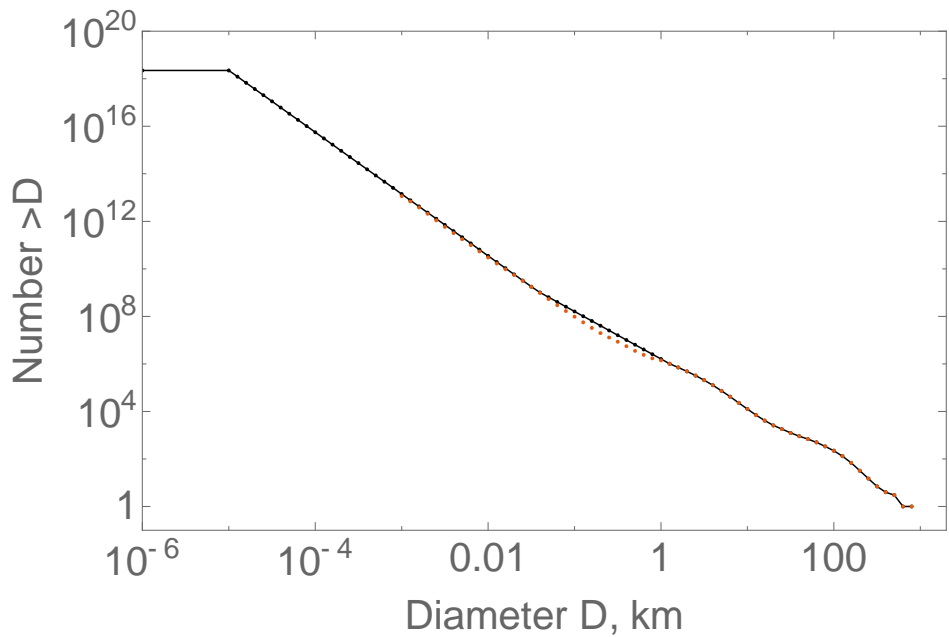

**Figure 1.** The assumed SFD and its binned representation for the main belt population. The red dot symbols are from Fig. 1 of Bottke et al., 2020, the curve labeled #1. The black curve, the KAH-SFD, is the same, but removes most of the dip at 0.1 km.

The uncertainty about the SFD for asteroids smaller than 1 km has a significant effect on studies involving small asteroids, such as surface lifetimes of small asteroids. Very small changes in the slopes assumed for 100 m asteroids will have a large numerical effect on the extrapolated numbers of sub-meter sized impactors. On the other hand, many of the results here do not depend upon that part of the population. *SSAH* does allow any population SFD to be read in from a user pre-defined data file, thereby facilitating studies about its effects.

### *2.2. The Shapes*

Asteroids have a wide variety of shapes, and those affect the mechanics of impacts. They are often assumed to be ellipsoidal, either prolate or oblate, with the ratio of longest (*a*) to shortest principal axis (*b*) ranging from *a/b*=1 for a sphere, to *a/b*=3 or more for the most elongated objects.

The light curve data in (Warner et al., 2019) for almost 5000 asteroids produces the following curve for the aspect ratio versus the diameter. (See also Mommert et al., 2018 for discussions on asteroid shapes).

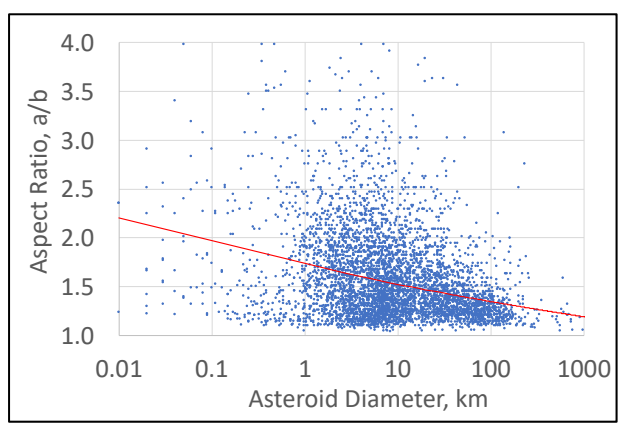

**Figure 2.** The estimated aspect ratios of about 5000 asteroids from light curve amplitudes as a function of their diameter, assuming prolate ellipsoidal shapes. The red line is a power law fit with the slope of -0.053. This is derived from Warner et al., 2019 using the maximum amplitude A and $\frac{a}{b} = 10^{\frac{A}{2.5}}$ .

The shapes and their evolution of those shapes is complex. In the simulations presented here, it is not possible to model this detail nor to follow shape changes from impacts. Instead, I include only the crudest measures, allowing shapes to become more elongated when excess spin requires shape changes. Improved models would be warranted.

*2.3. The Spins*

Knowledge of the spin magnitudes versus diameters of the asteroids has improved considerably in the last few decades. Binzel, 1979, listed the spins of 125 asteroids. Here I use the August 2019 compilation from the Light Curve Database (Warner et al., 2019) with over 17,000 asteroids in the main belt and NEO space, with the quality rating of '2' or better. I do not include Trojans, Hildas, Centaurs, or the somewhat controversial (e.g. Harris et al., 2012, Warner et al., 2020) main belt data from recent Palomar wide-field surveys.

The **Fig. 3** displays the data for the magnitude of the spin versus the object's diameter. The notes indicate strength and gravity spin limits, and the loci of constant tumbling recovery times, and other features which are discussed below. The suspected binary objects are from Pravec and Harris, 2007 and the tumblers are from Pravec et al., 2005.

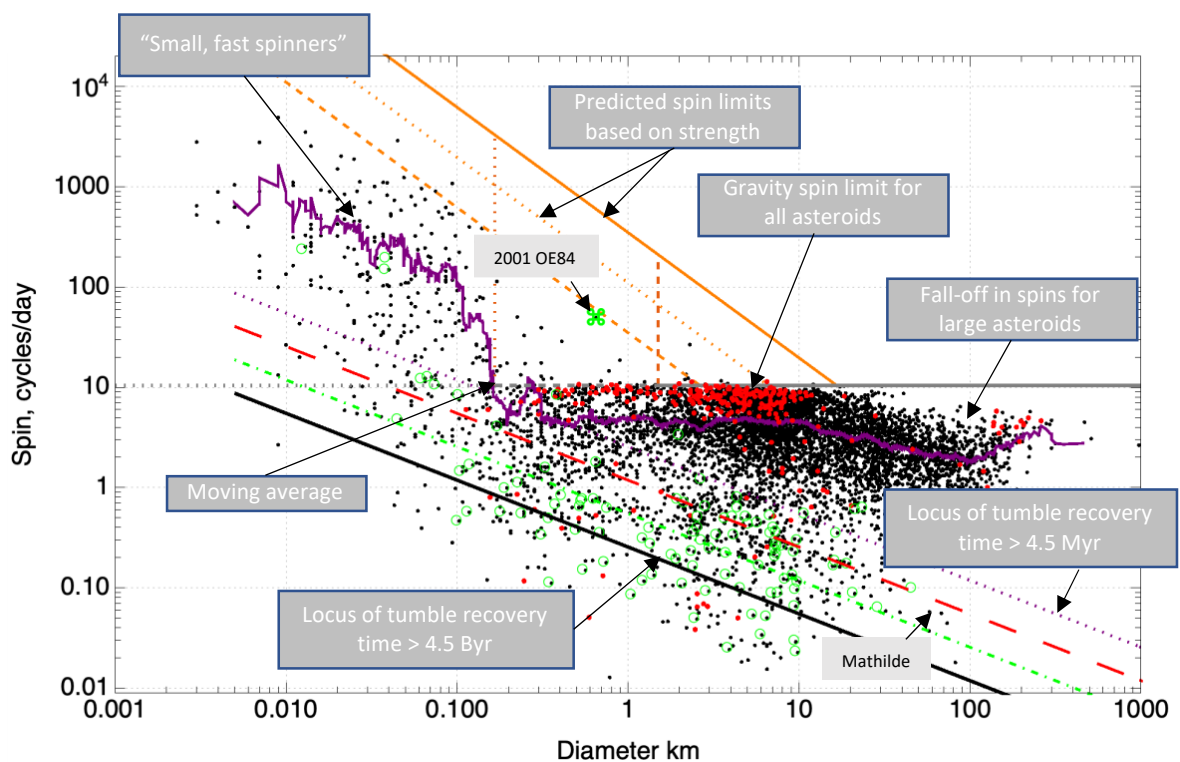

**Figure 3**. The spins versus diameters of about 17,000 observed asteroids in the main belt and NEO space (Warner et al., 2019). The known and suspected binaries, mostly grouped just under the gravity limit, have the red dots, they are less than 2% of the total data. The tumblers, about 3.5% of the total, have the green symbols. The "unusual" 2001 OE84 has the green clover-leaf symbol. The slow spinning, tumbling asteroid Mathilde is down towards the lower right. The orange curves are the strength spin limits for three values of strength, the upper one is for the expected strength, the lower one for 1% of that (Holsapple, 2007). Those will not be of any consequence here, as discussed below in §6.3.4. Other important features are identified on the plot.

This data is used below as the standard for outcomes of the simulations. However, a caution is in order. The asteroids with measured spins are only a small part of the entire set of asteroids, especially for those smaller than about 20 km, so this data is incomplete. There are known biases.

Dispersions about the average of this spin data are also informative. For a defined diameter range, a histogram of the spins shows that property. For the diameter range of 5 to 10 km, a histogram of the data is shown as **Fig. 4**:

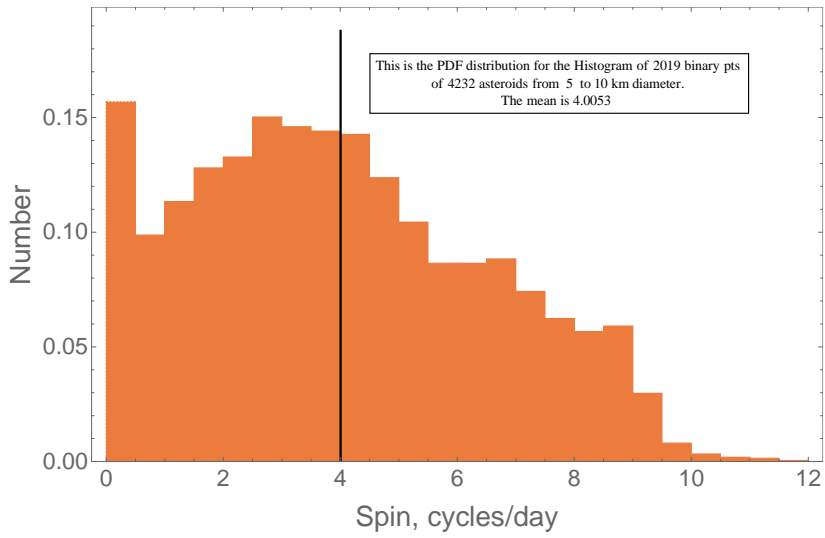

**Figure 4**. A histogram of the spins of the about 4000 asteroids with a diameter of 5 to 10 km. The gravity limit at about 10 cycles/day is evident. Below that, there is a broad distribution with a peak at about 4 cycles /day and a separate group of slowly spinning objects.

## *2.4. The Intrinsic Collision Probability*

The number and sizes of impactors that will strike a given target body in a time interval are determined by three items. The first is the population size-frequency distribution $N_{>D}$ discussed above. The second is the average spacing of those objects, which is determined by the volume they occupy. And the third are their relative velocities.

If $W$ denotes the total volume of the main belt, and $N_{>d}$ the number of asteroids with diameters greater than $d$, and if the asteroids are uniformly distributed, then the average number *per unit volume* of objects with diameter $> d$ is given as $N_{>d}/W$. In a time $t$, an asteroid of radius $R=D/2$ moving at a speed $U$ will sweep out a volume $\pi R^2 U t$ which, on average, will contain $\pi R^2 U t (N_{>d})/W$ objects with diameter $> d$. Opik, 1951, and Wetherill, 1967 defined the *intrinsic collision probability* $P_i = \pi U/W$, so that for a given asteroid with radius $R$, the expected number of hits $N^h_{>d}$ with diameter $>d$ of an asteroid in a time $t$ is determined from the total number of asteroids with diameter $>d$ as:

$$N^h_{>d} = P_i R^2 N_{>d} t. \qquad (2)$$

In this form, it is the intrinsic probability $P_i$ rather than $W$ that defines the spatial density of the asteroids.

That intrinsic probability can be estimated by studies of the trajectories of actual asteroids. Wetherill, 1967 made such a calculation and Farinella and Davis, 1992 presented a later one for different locations in the asteroid belt. They obtain the representative value $P_i=2.85 \times 10^{-18}/(km^2 yr)$, which is used here. There is no attempt here to model different probabilities in different parts of the asteroid belt, although that would be possible if target asteroids had an identified region in the belt.

## *2.5. Impact velocities*

In the belt, there is also a range of probable impact velocities. The probability density function (PDF) determined by Farinella and Davis, 1992 was used:

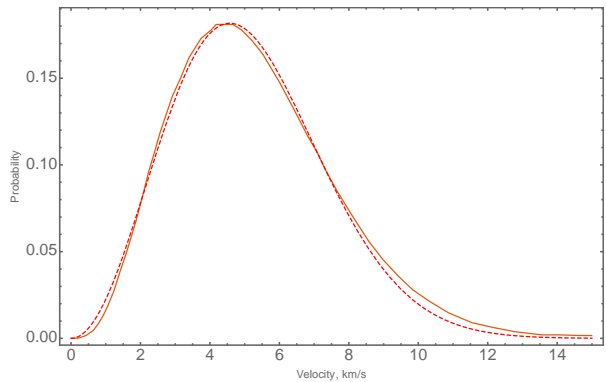

**Figure 5**. The distribution of asteroid impact velocities obtained by an integration of trajectories of all pairs of 682 main belt asteroids with diameter > 50 km. The most probable impact is at 4.4 km/s and the mean impact is at 5.29 km/s. The red dashed curve is my Maxwellian fit. *(From Farinella and Davis, 1992)*.

## *2.6. Spectral Types and Compositions*

The main belt has a multitude of asteroid spectral types. While those undoubtedly have a wide range of physical properties, little is known of those. That precludes a complete accounting for the differences.

In *SSAH* the two main types are assumed. The *C-Types*, the dark carbonaceous objects, are assumed to account overall for about 75% of the known asteroids; and the *S-Types*, stony (silicaceous) objects, to account for about 17%. The additional 8% are grouped with the S-Types here, although it would not be hard to include 8% of *M-types* and even others in the code if reasonable estimates of their properties were available. In *SSAH* the type assigned to a given target can be user-selected or can be assigned randomly using 75-25% percentages. There is no accounting for different mixes in different parts of the asteroid belt.

One of the most fundamental properties of the impact physics of an asteroid is its mass density, and the dependence of the scaling coefficient $\mu$ used in the point-source impact scaling (Holsapple, 1993) on that density. Unfortunately, for mass densities we only have data for a small number of objects. Here I use the Carry, 2012 data to support the estimates for an average density of 1.5 g/cm$^3$ for most C-Types with diameters <200 km) and 2.5 g/cm$^3$ for S-Types. Also, assuming the 75%-25% split, an average density over the entire belt would be 1.75 g/cm$^3$.

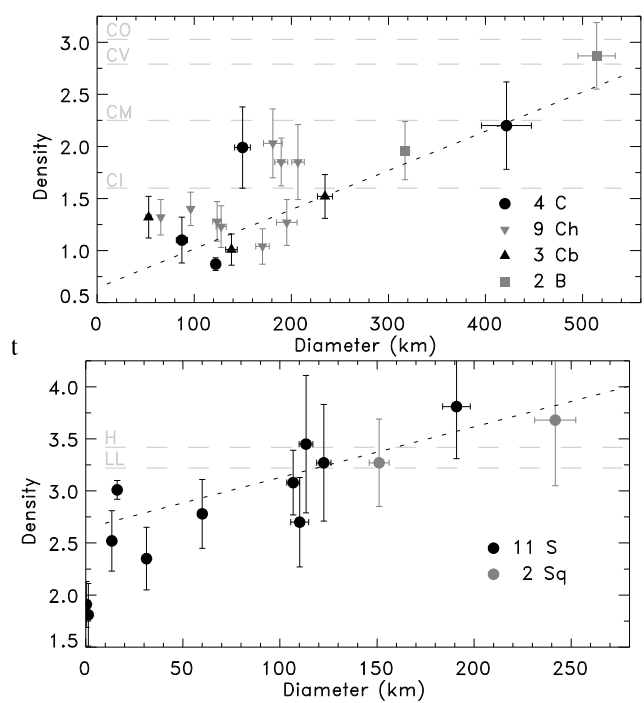

**Figure 6**. The global densities of C-complex asteroids (upper frame) and S-complex asteroids (lower frame). Here it is assumed that 75% of main belt asteroids are C-type, and, if their diameter<200 km, are relatively porous on a global scale with a density of 1.5 g/cm$^3$ with the scaling exponent $\mu=0.41$. The S-types are assumed for the remainder 25%, and they are all assumed to have little global porosity, so for any impactor larger than a few meters, it is assumed that the density is 2.5 g/cm$^3$, and $\mu = 0.55$. Other possibilities are ignored. From Carry, 2012.

## 2.7 SEVERITY OF IMPACTS: Q* MODELS

Since the studies of Davis at al., 1989 and Farinella at al., 1992, it has been known that the major contributor to both size changes and spin increments for the asteroids in the main belt is not the myriad of small impacts, instead it is those that approach the sizes causing disruption and or dispersion. Therefore, the measure of the disruption and or dispersion threshold is a crucial part of any study of the asteroid histories.

That measure is commonly denoted[4] as $Q^*$: it is the kinetic energy of an impactor per unit target mass that will disrupt the target and disperse half of its mass (CD events). Plots of $Q^*$ versus the target size for an assumed velocity are known as the $Q^*$ curves. A review of the theories and models for the those curves has been given in Holsapple and Housen, 2019; in their Fig. 2 is a plot of over 15 $Q^*$ curve estimates from the literature. The theoretical estimates given by the present author are based on the "point-source", (PS) assumption[5], characterized by a scaling exponent commonly denoted as $\mu$. (Holsapple, 1993). A curve for each of the two composition classes was chosen from those as the most representative and are the assumed defaults here. **Fig. 7** presents those curves for the average impact angle of 45° and average impact velocity of 5.5 km/s. Those can be adjusted for other impact angles and velocities using the scaling for these CD curves.

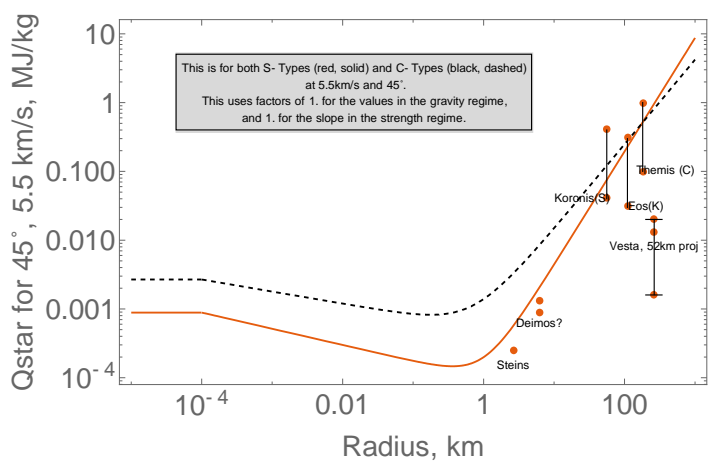

**Figure 7.** The assumed catastrophic dispersion limits for the two composition types for an impact at 5.5 km/s at 45°, as a function of the asteroid target radius. The points labeled 'Steins', 'Deimos' and 'Vesta' are from estimates of the specific energy which made their largest craters, which would be below that required for catastrophic dispersion. (See Holsapple and Housen, 2019).

Laboratory experiments in cm-sized samples set the left $Q^*$ intercept. Values corresponding to craters on Steins, Deimos and Vesta give a lower limit since those impacts did not disperse those objects. But the impact energy that created those large craters is not well known. The estimated energy for the Koronis, Themis and Eos families (Fujiwara, 1982) provide a possible range of values for the large objects, although that data is now old. The wide range results from a wide range of assumed percent of impact energy transferred to the orbital kinetic energy. The data for craters of Vesta are indicated because in some studies of asteroid histories its large craters are considered important constraints. They are well below the these assumed $Q^*$ dispersion criteria.

---

[4] I consider only disruption with dispersion, not disruption that reaccumulates. In the strength regime those two measures are the same but in the gravity regime the dispersion criterion is likely an order of magnitude higher than the disruption criterion.

[5] it is noted that the PS assumption and the resulting power law forms is a physical requirement for any process in which the input is in a very small region and very rapid compared to the entire domain. However, the value of the power law exponents is only determined to within limits: the appropriate value has been obtained from a large variety of experimental results.

There is little doubt that the low porosity C-Type objects require more energy per unit mass to disperse then do the higher density S-Types. That fact has been typically ignored in main belt studies to date. Holsapple and Housen, 2019 note that the curve for (assumed porous) C-Types is especially uncertain; the curve here is based entirely on code calculations by Jutzi, 2015.

Equations of these curves are expressed by two-part power-law functions in terms of the values *qc1* at an asteroid radius *R=10* cm in the strength regime (the first term), and a value *qc2* at *R*=250 km in the gravity regime (the second term):

$$Q^* = \left[ Min\left[ qc1 \left(\frac{R}{0.5\,10^{-4}}\right)^{3\mu\left(\frac{1}{1-2nsize}\right)}, qc1 \right] + \left[ qfac\, qc2 \left(\frac{R}{250}\right)^{3\mu} \right] \right] \left[\frac{UCos(\phi)}{5.5}\right]^{2-3\mu}. \quad (3)$$

The variable *nsize* defines the assumed decrease of strength with size as $Y \sim D^{1/nsize}$. The factor *qfac* is included in the code to allow for easy consideration of magnitude changes in the gravity regime. The dependence on the impact angle is discussed in §2.7.1 below. Note that these curves do depend on the impact velocity and angle.

It is assumed that $Q^*$ is constant below 10 cm target size. Above that, these, and all other current $Q^*$ estimates have common features. For targets from 10 cm up to several hundred meters or more, the $Q^*$ value is defined by a power law, decreasing with increasing target size. This is the "strength regime" which is governed by a material strength of the target. The decreasing value is a physical signature of the assumption that the material strength is decreasing with increasing event size, although that dependence is not well known[6].

For targets increasingly larger than a km or so, the curves turn upwards to a different power-law. This is the "gravity regime" where material strength is irrelevant and gravity controls impact events. The positive slope is a result of increasing self-gravity which holds the material together, creates internal compressive stresses, and makes it harder for mass to escape. In terms of the scaling exponent *μ* defining the point source scaling (e.g. Holsapple, 1993), that slope is *3μ*. Therefore, using the standard value *μ~0.55* for non-porous asteroids, that slope is about 1.65. For porous objects, *μ~0.4* so that slope is shallower, about 1.2. Physical limits restrict that slope from 1.0 to 2.0.

The two power-laws determine a composite curve that is, with a smooth transition, defined by 4 primary measures: a magnitude and slope in the strength regime and a magnitude and slope in the gravity regime. Two of those are reasonably well constrained by the physical requirements of point source processes, but two are not. The interested reader can refer to Holsapple and Housen, 2019. There it is noted that the slope in the strength regime depends upon how strength depends on asteroid size, which is not really known for asteroids. In the gravity regime, the slope is constrained by the point-source scaling laws, but the magnitude is not. The magnitude of the values in the gravity regime are unknown to at least factors of several. For those reasons, there are 'knobs'[7] in *SSAH* allowing changes in those two quantities for study.

---

[6] The decreasing material strength is attributed to a Weibull distribution of internal flaws in the target. Physical constraints limit that negative slope to a range from -.18 to -.33.

A problematic feature that should be noted are the values for the very largest asteroid targets. When the impactor and target have the same mass density, the ratio of the diameter $d^*$ of the impactor that disperses the target to that of the target is given by the expression $d^*/D = (2Q^*/(UCos(\phi)^2))^{1/3}$. For a large, say 300 km target, the above curve has $Q^* > 1$ MJ/kg. So, assuming an average impact speed of 5.3 km/s, gives $d^*/D = 0.5$. And the curve value for a 1000 km targets predicts a diameter ratio $d^*/D = 1.1$; an impactor larger than the target! That creates a conundrum; that the extrapolation of the scaling slope to the largest diameters gives impactor sizes that contradict the original point-source premise that they are small compared to the target. It may be that a decrease in the $Q^*$ curve at the upper end would be appropriate, although that issue is not addressed further here. It is only an issue for the largest asteroids.

### *2.7.1. Impacts at an Angle*

Most lab data, calculations and theories are for impacts perpendicular to a target surface. However, over 75% of impacts in the belt occur at angles between 20˚ and 70˚ with an average of 45˚ (§3.2.1). Since experiments by Gault and Wedekind, 1978, it has been known that craters are essentially round for all impact angles above about 20˚ from the horizontal. Further, existing data for final outcomes of an impact process are generally consistent with the simple assumption (Chapman and McKinnon, 1986) that an impact at angle $\phi$ from the vertical and with velocity $U$ has the same cratering and mass removal results as a normal (0˚) impact with the reduced velocity $UCos(\phi)$.

That simple assumption is made here for all cratering and mass removal models. But note that the imparted spin depends crucially on that impact angle, see §3.2. I will also consider possible interaction between impacts at an angle and pre-existing spin, in the section about angular momentum drain in §3.2.3.

## 2.8. HISTORIES USING AVERAGE INPUTS

This paper develops a simulation approach that calculates explicit impact results for each large impact; but approximate results using average input values are useful for guidance and checks[8]. And, as noted above, such average calculations were the basis for many previous approaches. Here I present some results using average inputs for the population, intrinsic probability, dispersion curve and impact conditions, assuming a total time of 4.5 Byr[9]. These are based on an object's initial diameter and ignore the decrease of its diameter with time. Especially for a small asteroid, that erosion can be very important (see §7.4).

---

[7] The colloquial term 'knob' is used for means to adjust poorly known code input parameter values to test their effects on output values. They can also be called "ignorance factors".
[8] Although one should recall that in a nonlinear system, the results using average inputs can be substantially different than the average result.
[9] It is not intended to imply that the population was in fact steady for that entire history. That choice provides an important end case.

These results depend upon what one assumes for the impact angle. The average angle is 45˚, that is not often used in their literature for calculations of this type. To be more consistent with those previous calculations, here I assume an impact perpendicular to the surface. Since the average angle is 45 degrees, these estimates are affected by that choice.

It should be noted that these results have a large error band. Consider for example the mean lifetime for a 1 km asteroid. The diameter of the impactor required to catastrophically disperse that asteroid is about 40 meters. The lifetime estimate depends inversely on the number of assumed objects smaller than that diameter. The SFD possibilities for those small objects can easily vary by a factor of several or more. For that reason, any result based upon the SFD for meter and smaller sized asteroids cannot be stated with any reliability.

Some results are provided in the following Table 1 for the two asteroid types. The subsequent paragraphs discuss how those estimates were calculated.

| S-Type | | | | | | | | |
|---|---|---|---|---|---|---|---|---|
| Target diameter, km | Mean largest impactor in 4.5 Byr, km, dL | Q* (MJ/kg) | Dia d* to bust target, km | Average Lifetime, yr | Prob of being dispersed or pulved in 4.5 Byr | Smallest significant impactor, km | Number of Significant hits | Time between significant hits, Myr |
| 0.0001 | 5.2E-5 | 1.E-3 | 5.E-6 | 26.E+6 | 100 | 4.7E-7 | 469.E+0 | 6.40E+3 |
| 0.001 | 3.0E-4 | 684.E-6 | 41.E-6 | 47.E+6 | 100 | 4.1E-6 | 25.E+3 | 1.20E+2 |
| 0.01 | 1.8E-3 | 398.E-6 | 344.E-6 | 118.E+6 | 100 | 3.4E-5 | 28.E+3 | 1.05E+2 |
| 0.1 | 1.0E-2 | 232.E-6 | 3.E-3 | 295.E+6 | 100 | 2.9E-4 | 11.E+3 | 2.64E+2 |
| 1 | 7.2E-2 | 170.E-6 | 26.E-3 | 885.E+6 | 99 | 2.6E-3 | 4.E+3 | 7.91E+2 |
| 3 | 2.2E-1 | 319.E-6 | 96.E-3 | 1.E+9 | 96 | 9.6E-3 | 1.E+3 | 2.67E+3 |
| 10 | 7.2E-1 | 2.E-3 | 552.E-3 | 4.E+9 | 64 | 5.5E-2 | 168.E+0 | 1.79E+4 |
| 30 | 2.4E+0 | 10.E-3 | 3.0 | 11.E+9 | 33 | 3.0E-1 | 52.E+0 | 5.81E+4 |
| 100 | 7.1E+0 | 70.E-3 | 19.3 | 76.E+9 | 6 | 1.9E+0 | 16.E+0 | 1.82E+5 |
| 300 | 1.7E+1 | 431.E-3 | 105.9 | 142.E+9 | 3 | 1.1E+1 | 3.E+0 | 9.61E+5 |
| 1000 | 8.3E+1 | 3.E+0 | 684.3 | 702.E+9 | 0.6 | 6.8E+1 | 1.E+0 | 2.13E+6 |

**Table 1a**. Representative features of the histories of S-Type asteroids of various diameters. These results assume average impactor density, impact velocity, and angles over a 4.5 Byr history. The erosion of size over that period is neglected.

| C-Type | | | | | | | | |
|---|---|---|---|---|---|---|---|---|
| Target diameter, km | Mean largest impactor in 4.5 Byr, km, dL | Q* (MJ/kg) | Dia d* to bust target, km | Average Lifetime, yr | Prob of being dispersed or pulved in 4.5 Byr | Smallest significant impactor, km | Number of Significant hits | Time between significant hits, Myr |
| 0.0001 | 5.2E-5 | 2.00E-3 | 5.E-6 | 2.0E+7 | 100 | 5.0E-7 | 466.E+0 | 6.4E+3 |
| 0.001 | 3.0E-4 | 1.51E-3 | 45.E-6 | 2.5E+7 | 100 | 4.5E-6 | 23.E+3 | 1.3E+2 |
| 0.01 | 1.8E-3 | 1.01E-3 | 395.E-6 | 6.3E+7 | 100 | 4.0E-5 | 20.E+3 | 1.5E+2 |
| 0.1 | 1.0E-2 | 6.85E-4 | 3.E-3 | 1.6E+8 | 100 | 3.5E-4 | 7.E+3 | 4.3E+2 |
| 1 | 0.072 | 6.87E-4 | 35.E-3 | 4.7E+8 | 100 | 3.5E-3 | 1738.4 | 1.7E+3 |
| 3 | 0.216 | 1.29E-3 | 129.E-3 | 9.0E+8 | 99 | 1.3E-2 | 525.4 | 5.7E+3 |
| 10 | 0.718 | 4.37E-3 | 644.E-3 | 2.7E+9 | 81 | 6.4E-2 | 124.1 | 2.4E+4 |
| 30 | 2.4 | 1.60E-2 | 3.0 | 6.8E+9 | 49 | 3.0E-1 | 52.0 | 5.8E+4 |
| 100 | 7.1 | 6.93E-2 | 16.2 | 5.2E+10 | 8 | 1.6E+0 | 22.0 | 1.4E+5 |
| 300 | 17.3 | 2.67E-1 | 76.1 | 7.9E+10 | 6 | 7.6E+0 | 7.4 | 4.1E+5 |
| 1000 | 83.0 | 1.17E+0 | 415.7 | 1.4E+12 | 0 | 4.2E+1 | 2.8 | 1.1E+6 |

**Table 1b**. Representative features of the histories of C-Type asteroids of various diameters. These results assume average impactor density, impact velocity, and angles with a 4.5 Byr history. The erosion of size over that period is neglected.

### 2.8.1. Largest Impactor Expected in a Time Interval

The mean largest impactor diameter $d_L$ to hit any single asteroid with diameter $D$ in a time $T$ is found by setting the expected number of hits in Eq. (2) to unity, giving the number of such objects with $d > d_L$ as $N_{>d_L} = 1/(P_i R^2 T)$. Then the diameter $d_L$ is found by solving for $d_L$ by inverting (perhaps numerically) the SFD function for the population. The quantitative results for the small objects can vary significantly depending on the choice of the population. Numerical results for the two asteroid types are presented in the 2$^{nd}$ column of Table 1.

Using the present SFD of **Fig. 1**, over the main belt entire lifetime of 4.5 Byr, a 10 km diameter target of either type has the largest expected impactor diameter of about 0.7 km. A 100 km diameter target has the largest expected impactor of about 7.1 km.

### 2.8.2. Percent of Targets Catastrophically Dispersed in 4.5 Byr

From the largest expected impactor data, statistics on the probability of a catastrophic dispersion (CD) can be obtained. Let $d^*$ be the impactor mass that would CD some given target asteroid and $d_L$ the expected largest impactor in each time interval, as given above. Using Poisson statistics, the probability of a hit with the diameter $d^*$ or larger in that time is given by

$$prob(d > d^*) = 1 - Exp\left[-\left(\frac{N_{>d^*}}{N_{>dL}}\right)\right]. \quad (4)$$

The values for $Q^*$ for the target sizes are listed in columns 3 and the corresponding diameters $d^*$ required for a catastrophic disruption are listed in column 4. Again, the SFD is used to determine the numbers of impactors. Then the probability of a catastrophic impact in the 4.5 billion years is tabulated in column 6.

A 10 km diameter C-Type asteroid requires a $d^*$=600 m (at 0°) impactor to disperse it. Thus, the probability in 4.5 Byr of getting hit by that or a larger impactor is about 80%. From an initial population of 100 asteroids of that size, only 20% would survive a 4.5 Byr history.

For a 100 km S-Type, the probability in 4.5 Byr of a 19 km impact normal to the surface required to bust it is only 6%, thus it is 94% likely to survive that time. The fragments form the dispersed 6% could be a source of asteroid families with the largest body having a diameter some fraction of 100 km. One should note that this average analysis is misleading for small objects, because of the failure to allow for target erosion; better estimates will emerge from the simulations. And the estimate for families is discussed in §7.3.

Results for other target sizes are also given in Table 1. An asteroid with a diameter less than a few km has little chance of surviving 4.5 Byr. Therefore, all sub-km asteroids are remnants from the breakup of a larger object. And, at the other extreme, for the dozen largest asteroids there is negligible chance of an impact large enough to disperse it in 4.5 Byr.

### 2.8.3. Significant Impacts

Every target body will be impacted by a multitude of small impactors over its lifetime, but the smallest impacts will have very small effects. The calculations in *SSAH* are separated into two categories. For the larger ones, explicit, detailed case-by-case calculations are made. For the smaller "insignificant" ones, averaged implicit calculations are made.

Arbitrarily, here a 'significant' impactor is defined as one with a kinetic energy per unit target mass greater than $10^{-3}$ of the value $Q^*$ that would catastrophically disrupt the target object. Therefore, the largest significant impactor has a diameter $d_{min}$ that is 10% of the diameter $d^*$ of the impactor that would disperse the target. That significance limit diameter $d_{min}$ can be calculated using the average impact velocity of 5.29 km/s and a 45° impact angle and for a given population. Then the number of impacts expected that are larger than that size is given by

$$N^h_{>d_{min}} = (P_i R^2 t) N_{>d_{min}}. \tag{5}$$

Those are also tabulated in the Table 1. The larger bodies have many fewer significant hits compared to the smaller ones: for example, a 100 km C-Type asteroid has only about 17 such impacts in 4.5 Byr, while a 1 m one has 20 thousand.

### 2.8.4. Remaining Lifetimes

Since every small asteroid is a remnant from the destruction of a larger ones, I use the term "remaining lifetimes" for the expected time before it is catastrophically dispersed itself, given a present state. A first order estimate of the expected remaining lifetime of an asteroid can be found here using the approach common in the literature: the average time interval before a single large impact occurs with impactor energy per target mass greater than the threshold $Q^*$. That would have its mass is reduced to less than half of the original. The difference between an original diameter and the depleted diameter at the time of catastrophic dispersion is ignored, although small objects will have eroded significantly before a CD event (as noted by Dohnanyi, 1969). Below, in §7.4 the lifetimes will be reconsidered using the detail of the simulations.

The $Q^*$ curve and a choice for the velocity $U$ defines the impactor diameter $d^*$ for a CD event for a target of radius $R$. I equate the largest expected $d_L$ to $d^*$ and solve for the average time interval for which an impactor of that size is expected for a target of radius $R$. The result, the average time for a catastrophic event assuming the present main belt conditions, is given as

$$T(D) = (P_i R^2 N_{>d^*})^{-1}. \tag{6}$$

This time depends significantly on both the $Q^*$ curve and the population via of the $N_{>d^*}$ term. Columns 5 give the estimated remaining lifetimes of the various targets. Corresponding to the dispersion probabilities, the m-sized objects have predicted lifetime of less than 100 Myr, while asteroids above 100 km diameter have predicted lifetimes many times that of the solar system.

The lifetime for any given asteroid scales inversely with the number in the population of objects greater than a fixed diameter. Thus, the estimate of the lifetime of a small asteroid can change by several orders of magnitude using different populations. For all except for the largest asteroids, that fact can make lifetime estimates highly variable. Until more is known of the population of small impactors, we cannot make accurate time-scale estimates for small asteroids.

## 3. MODELING OF IMPACT EVENTS

The process of an impactor colliding with a target object is a very complicated time-dependent one, occurring mostly over a time interval on the order of a few seconds per km target diameter. During that interval, shock waves are traveling back and forth in the target, material is being compressed, moved, launched, and cast permanently from the target. That results in a dynamical phase with individual particles moving around under gravitational forces; and some will re-accumulate with the original target, or perhaps in separate identifiable but bound masses such as satellites. Ultimately the target and impactor are left in a new state with a different shape, a different mass, and a different spin.

Each of the resultant objects has a mass magnitude, a spin, and an angular momentum measured around the original center of mass of the target. Those of the target are calculated here. There are no calculations of the motions of fragmented particles in space.

### *3.1. Mass Loss Models*

At the high relative velocities (~5 km/s) of the present main belt, every impact is strongly erosional, removing 100 to 1000 times the mass of the impactor, depending on the impact severity. In an angled impact, mass first launched downstream from the surface is a luminous high-speed stream emanating from the impact point. This probably includes less than 1% of the projectile mass , and a much smaller percent of the total ejecta (Yanagisawa and Hasegawa, 2000a). The main ejecta is launched next: it is a dense ejection of target material. For that main ejecta, for angled impacts, Schultz, 1999 noted asymmetries at very early times (lower angle downrange ejecta, higher angle up-range ejecta) but it soon becomes symmetric in a cone with axis perpendicular to the local surface. That is how the majority of mass (and momentum) is ejected.

The target mass loss is the amount of ejecta that escapes the target's gravitational field, not any part that falls back to the target surface. The amount of ejecta which escapes the target body depends upon the gravity (size) of that target body. As the impact severity increases to that for catastrophic dispersion (CD), the mass loss will approach its 50% defined value; substantially greater impacts will pulverize and scatter the target into many small pieces.

The models for these processes are based on the impact scaling theories (Holsapple, 1983, Holsapple and Schmidt, 1987, Holsapple, 1993) for cratering, ejecta and catastrophic dispersions. The cornerstone of that scaling is the simplification allowed by the concept of a "point-source" (PS) measure for the impactor. That impactor measure governs the outcome because the initial effects of an impact on a target occur in a very small region compared to the size of the final crater; and almost instantaneously compared to the time extent of the evolution of the cratering event. Models based on that PS scaling have been constructed for cratering, ejecta and CD.

### *3.1.2. Crater Sizes*

For impact cratering, every size measure of the crater: its mass, volume *V*, or diameter *d*, depends on the PS measure of the impactor, and on whether some material strength *Y* (strength regime, small events) or the target surface gravity *g* (gravity regime, larger events) controls the excavation. The variables are commonly gathered into dimensionless groups:

$$\pi_V = \frac{\rho V}{m}, \quad \pi_2 = \left(\frac{ga}{U^2}\right), \quad \pi_3 = \frac{Y}{\rho U^2}. \tag{7}$$

In the strength regime (see Holsapple, 1993), ignoring small mass density terms, the crater volume *V* is a power of the strength group $\pi_3$:

$$\pi_V = K_2 (\pi_3)^{\frac{-3\mu}{2}}, \tag{8}$$

while in the gravity regime it is a power of the gravity from

$$\pi_V = K_1 (\pi_2)^{\frac{-3\mu}{3+\mu}}. \tag{9}$$

In the evaluation of these equations, the strength *Y* is assumed to depend upon the target diameter *D* according to

$$Y = Y_0 \left(\frac{D}{D_0}\right)^{\frac{1}{n}}. \tag{10}$$

This feature of the scaling is uncertain: in principle it is present, but its magnitude does not really have much supporting data. The value for the exponent *n* is taken here as 4, but it is uncertain. That value will have a major effect on simulations and life estimates of asteroid surface cratering.

The scaling exponent $\mu$ and the two constants $K_1$ and $K_2$ are found from a body of cratering results, both for explosives and for impacts. From the volume, the crater radius *R* is related to the volume *V* by

$$R = K_R V^{\frac{1}{3}}, \tag{11}$$

where $K_R$ is determined by the crater type. For a bowl-shaped crater it has a typical value of 1.3. This radius is adjusted for spall craters and for complex craters.

A discussion of the theory, definitions of all the variables, and plots of scaling are available from links at keith.aa.washington.edu/cratering. That site gives cratering results, gathered from many explosive and impact experiments for a range of geology types, to estimate the scaling constants for asteroids. In *SSAH* those constants are given the representative values $\mu=0.4$, $K_1=0.15$ and $K_2=1$ for *C-Types*, and $\mu=0.55$, $K_1=0.05$ and $K_2=1$ for *S-Types*.

Example curves, for a 100 km S-Type target with size-dependent strength and using various impact velocities are reproduced in **Fig. 8**:

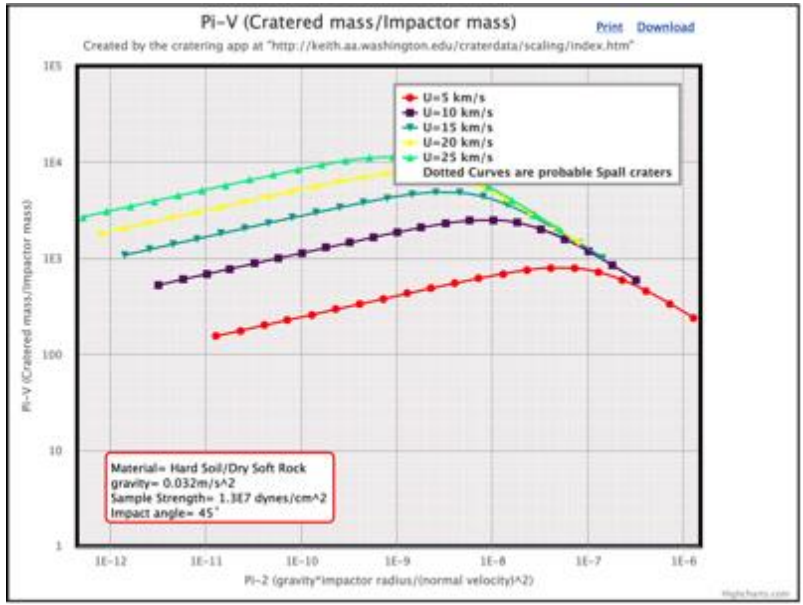

**Figure 8.** The scaled crater mass for a 100 km asteroid as a function of the impactor parameters. The curves at the left half are in the strength regime with size-varying strength with the exponent *n=4*. Those curves coalesce into a single downward sloping curve at the right, that is the gravity regime. The impactor diameter and target gravity determine $\pi_2$, then $\pi_V$ determines the crater size and, together with the ejecta velocity scaling, the mass lost in an impact.

The results for simple bowl-shaped craters suffice for the large majority of calculations of mass loss. But other crater shapes also occur. Small impacts in brittle targets (*S-Types*) can create a spall crater (Holsapple and Housen, 2013) which can be important for cratering on small asteroids; and large craters in solar system bodies (not asteroids) can have late-stage readjustments that morph the crater into a complex crater. Simulations in *SSAH* that include tracking individual craters include both of those possibilities. See §7.2 about spall craters on the asteroid Gaspra.

### *3.1.3. Escaping Ejecta*

Mass loss in a cratering event depends on the amount of the ejecta that escapes from the target. The ejecta characteristics can be described by a plot of ejecta mass with velocity greater than some velocity *v* versus that velocity *v*. Such a curve, using the PS scaling theory is shown in **Fig. 9** (from Housen and Holsapple, 2011). The ejected mass velocity varies from a largest $v_L$ launched from the crater center to a slowest *v\** launched from the crater rim. Between those values, based on the PS scaling, the mass with velocity greater than the velocity *v* is given as

$$M(>v) = M^* \left(\frac{v}{v*}\right)^{(-3\mu)}. \tag{12}$$

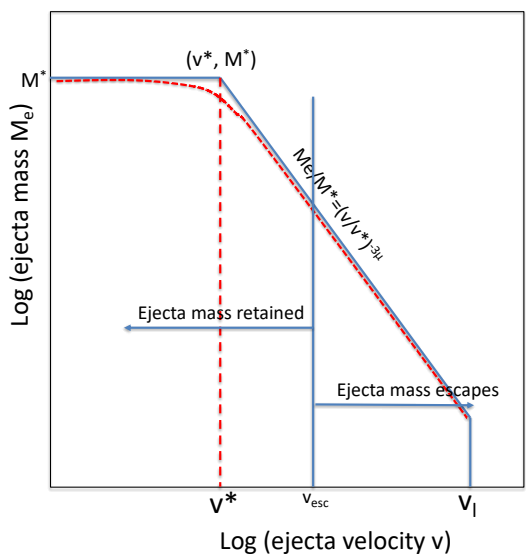

**Figure 9.** The amount of mass with ejecta velocity greater than some value *v*. The primary sloped portion is a power-law decreasing with increasing velocity from the coordinates *(v\*, M\*)* down to the minimum velocity $v_L$. All of the mass has a velocity greater than *v\**, thus *M\** is the total ejecta mass. All ejecta with a velocity greater than the escape velocity $v_{esc}$ will escape from the target body; that with smaller velocity will fall back to the target surface and will not contribute to mass loss nor spin-up. The escape velocity depends upon the target size and can be smaller than *v\**, in which case all ejecta escapes, or larger than $v_L$ in which case no ejecta escapes.

*M\** is the total ejecta mass which, according to the PS theory, is always some fixed proportion of the crater mass; the remainder of the crater volume is due to surface uplift and mass movement. That ejecta mass can be estimated as about 60% of the crater mass[10]. The value of the slowest ejecta velocity $v^*$ is found from laboratory experimental results for ejecta (e.g. Housen and Holsapple, 2011). For asteroids, it has the values given in Holsapple and Housen 2012, Eqs. (27):

$$v^* = K_{vs}\sqrt{\frac{Y}{\rho}}, \quad v^* = K_{vg}U(\pi_2)^{\frac{1}{2+\mu}}, \tag{13 a, b}$$

in the strength and gravity regimes, respectively. These two additional ejecta scaling constants are then found from experimental data. *SSAH* uses $K_{vs}$ and $K_{vg}$ =0.3 for both composition types (Holsapple and Housen, 2012). These equations and the value for the escape velocity define the amount of crater ejecta that will escape an asteroid.

### *3.1.4. Catastrophic Disruptions and Pulverizations*

The cratering and ejecta theory above is for craters in a half-space, with no influence of the finite target radius. But those are not the primary mass removal events in an asteroid's history. A large impact can create a crater with a diameter on the same order as the target diameter, and target curvature effects become important. The amount of mass removed from the target is increased because of regions of mass that are "blown out" from the "wings" of the crater. Furthermore, for a large target in space, fragmented material can return to the target core during gravitational re-accumulation processes. That is the feature that distinguishes disruption of a target versus dispersion of a target.

By definition, CD impacts disperse (remove) one-half of the mass. In that case there is no need to separately consider crater mass and ejecta mass or even reaccumulated mass, it is inherent in the choice for the CD curve. But what is the amount of mass removal for impact larger than the energy of the CD limit?

While not plentiful, there is some lab data on the mass loss from impacts approaching and exceeding the CD limit, all in the strength regime. **Fig. 10** (from Holsapple et al., 2002) is a plot of the largest remaining mass versus the impactor energy per unit target mass *Q,* grouped according to fundamental target composition. Those impacts with *qratio=Q/Q\*>1* remove more than half the mass.

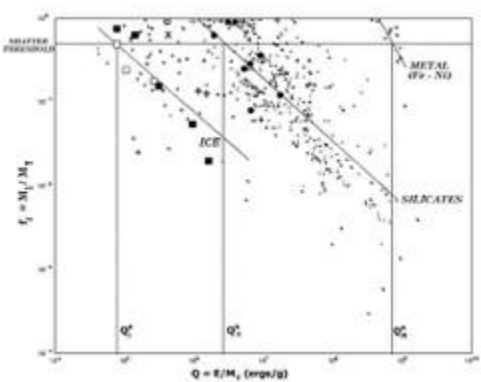

**Figure 10.** The mass loss for impacts less than, and greater than, the catastrophic dispersion limit (the $M_L/M_T$ =0.5 line), grouped according to basic composition. The crossover of the curves with that line define *Q\**. The slopes are about -2/3. (Holsapple et al., 2002)

---

[10] This directly connects the ejecta mass to the crater mass. The remainder is due to subsurface movement of material: downward in the crater center and upward in the crater rim and beyond. In Holsapple and Housen, 2012 the ejecta mass was found indirectly from measured ejecta results, not directly from crater mass. I now think this present simpler approach is best.

But these lab experiments only measure the initial fragmented mass. That creates fragments that fly away from the target. In space those become a collection of individual objects that gravitationally interact and reaccumulate into the final configurations. Many return to the remaining target mass. That process is not present in the lab experiments.

In the last few years, a few researchers have used code calculations to study the processes of catastrophic and super catastrophic impacts into large bodies. Often those codes use a two-part approach where the fragmentation phrase is calculated by a shock-physics code (several use an SPH code), and then fragments identified are passed off to an N-body code to calculate the resultant dynamics and re-accumulation. Those results might give guidance to the outcomes of impacts exceeding the CD threshold.

However, it is often true that in those codes, in the fragmentation modeling, the material is entirely broken down to the scale of the cells of the calculation. That is concerning and a non-physical feature. It suggests flaws in the strength modeling. Those codes need to be applied to the small laboratory experiments to verify and validate the fragmentation portion of the calculation. I am hesitant to use their results in this study.

Used here is an entirely empirical function to define the largest remaining mass when *qratio>0.5*, indicating a near-catastrophic, catastrophic, or super-catastrophic event. That function must have the value of 1/2 when *qratio* is equal to unity and must decreases rapidly towards zero for larger *qratio*. At a limit *qratio=pulvlim* the target is assumed to be pulverized. That variable *pulvlim* is then an input choice of the user.

That function, with those properties, is given as

$$M_R/M = \left[0.5 - \frac{0.5}{Log[pulvlim]} Log[qratio]\right]. \quad (14)$$

Those events with *qratio>pulvlim* are flagged as super-catastrophic, or *pulverizations*, and are no longer tracked.

### 3.2. Spin Increments and Accumulations

Every impact will add an increment of spin. Spin is represented by a three-dimensional spin vector **ω,** an angle per unit time. The spin increment depends on the impact location on the target surface, the impact angle, the impactor and target mass, and velocity, and by the amounts, directions, and speeds of the removed mass.

Lost mass carries away both mass and angular momentum (AM). That feature makes the analysis of spin more difficult than the analyses of mass loss. Different mass-loss mechanisms would have different amounts of angular momentum loss. For example, the early-time, down-range spray from low angle impacts would carry away AM. But, at least for a spherical target, the primary ejecta is cast away from the surface in a cone directed symmetrically about an axis pointing through the center of mass; so it so it does not contribute to AM loss. Any process with mass loss that occurs symmetrically about the impact direction, such as loss of mass around the equator, will not contribute to change of angular momentum.

A simple empirical approach to account for the net effect of these and other complications was first given by Burns and Safronov, 1973. They suppose that the AM retained in the target is some factor $\zeta(\phi)$ of the impactor AM, depending on the impact angle $\phi$. There is no accounting for the exact reasons for the final angular momentum of the target. In principle this factor could be found from experiments or perhaps high-fidelity code calculation.

That is the approach used for the simulations here. Experimental determination of that factor will be discussed, and then some considerations about how it might change in different situations. First the angular momentum of the impactor is modeled.

### *3.2.1. Geometry of a Single Impact: The Impactor Angular Momentum*

The impact geometry can be defined for a spherical target using two coordinate systems: a body-fixed global one and a local one. Those are illustrated[11] in **Fig. 11**. The current spin can be in any direction relative to the global axes. The impacts are assumed to occur randomly at all angles relative to the principal axes of an asteroid.

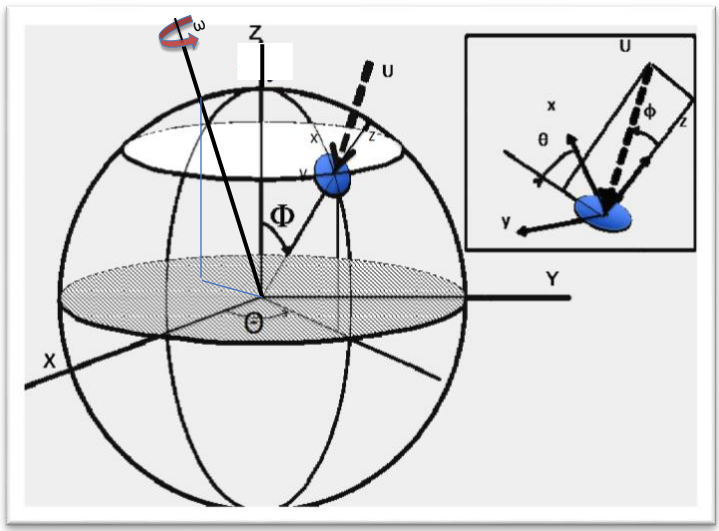

**Figure 11**. The global *X, Y, Z* Cartesian coordinates are aligned with the principal axes of inertia of the target body, with the *Z* direction having the largest principal angular inertia. The angle $\Theta$ is the azimuthal angle and $\Phi$ is the polar angle (colatitude) of the impact location in the corresponding spherical system. The inset shows a local *(x, y, z)* coordinate system at the impact point. The local *z* axis is normal to the surface, the local north or *x* axis is along a meridian, in the plane of the *Z* axis and the *z* axis, and the *y* axis is along a parallel. The impact has the spherical polar angle ø and azimuthal angle $\theta$ in that local coordinate system[12].

An impactor with the velocity vector $U$ strikes the target at a point on the surface of a spherical target body defined by random azimuthal and polar angles $\Theta$ and $\Phi$. The current spin vector $\omega$ has three components: $\omega = \{\omega_X \quad \omega_Y \quad \omega_Z\}$. If the current spin is not around the fixed *Z*-axis, it is said that the object is 'wobbling'. Severe wobbling is called 'tumbling'. Here I use the two words interchangeably.

The transformation between the unit base vectors of the local system and those of the global system is

$$\begin{bmatrix} e_X \\ e_Y \\ e_Z \end{bmatrix} = \begin{bmatrix} -Cos\Phi Cos\Theta & Sin\Theta & Sin\Phi Cos\Theta \\ -Cos\Phi Sin\Theta & -Cos\Theta & Sin\Phi Sin\Theta \\ Sin\Phi & 0 & Cos\Phi \end{bmatrix} \begin{bmatrix} e_x \\ e_y \\ e_z \end{bmatrix}, \qquad (15)$$

---

[11] Although the geometry is approximated as a spherical target, it is assumed that the principal direction Z is in a unique direction in the asteroid.
[12] The ø, $\theta$ and $\Phi$ here are denoted as $\theta$, –ø and $\Psi$ in Dobrovolskis and Burns, 1984, and as $\theta$, ø and $\Theta$ in Yanagisawa and Hasegawa, 2000).

with the inverse transformation

$$\begin{bmatrix} e_x \\ e_y \\ e_z \end{bmatrix} = \begin{bmatrix} -Cos\Phi Cos\Theta & -Cos\Phi Sin\Theta & Sin\Phi \\ Sin\Theta & -Cos\Theta & 0 \\ Sin\Phi Cos\Theta & Sin\Phi Sin\Theta & Cos\Phi \end{bmatrix} \begin{bmatrix} e_X \\ e_Y \\ e_Z \end{bmatrix}. \quad (16)$$

Consider a single impactor with mass *m* and velocity **U**, with the velocity magnitude *U*, impacting a target with radius *R*. The angular momentum (AM) of that impactor about the target asteroid center of mass is $\delta h_i = \delta m(r \times U)$, where *r* is the radius vector to the impact point. It is given in terms of the unit base vectors of the local coordinate system as

$$\boldsymbol{\delta h_i} = RmU Sin(\phi)[Sin(\theta)\boldsymbol{e_x} - Cos(\theta)\boldsymbol{e_y}], \quad (17)$$

where $\phi$ is the angle of the velocity vector from the surface normal. The magnitude of the angular momentum of the impactor is

$$|\boldsymbol{\delta h_i}| = RmU Sin(\phi). \quad (18)$$

In terms of the global axes, Eq. (17) gives the angular momentum of the impactor in the global coordinates as

$$\boldsymbol{\delta h_i} = |\boldsymbol{\delta h_i}| \begin{bmatrix} -(Sin(\theta)Cos(\Theta)Cos(\Phi) + Cos(\theta)Sin(\Theta)) \\ (-Sin(\theta)Sin(\Theta)Cos(\Phi) + Cos(\theta)Cos(\Theta)) \\ +(Sin(\theta)Sin(\Phi)) \end{bmatrix}^T \begin{bmatrix} e_X \\ e_Y \\ e_Z \end{bmatrix}. \quad (19)$$

Therefore, the angle $\alpha$ between the angular momentum increment vector and the vertical **Z** axis is given by

$$Cos(\alpha) = Sin(\theta) Sin(\Phi). \quad (20)$$

For a single impact, the probability distributions for the relevant angles can be calculated from the ratio of the differential of the area element exposed along the direction of the impact, compared to the entire surface area[13]:

$$f(\phi, \Phi, \Theta) d\phi d\Phi d\Theta = \frac{R^2(2 Cos(\phi) Sin(\phi) Sin(\Phi))}{(4\pi R^2)} d\phi d\Phi d\Theta \quad (21)$$

The angle $\Theta$ is uniformly distributed on $[0, 2\pi]$. One can integrate over its range and over $\Phi \to [0, \pi]$ to get the probability density for $\phi$ as $f(\phi) = 2 Cos(\phi) Sin(\phi) = Sin(2\phi)$, and then that for the angle $\Phi$ is $(\sin \Phi)/2$. Thus, the most probable impact angle is 45° and almost 75% of impacts occur at an angle between 20° and 70°. The cumulative distributions for $\phi$ and $\Phi$ are, respectively, $F(\phi) = (1 - Cos(2\phi))/2$ and $F(\Phi) = (1 - Cos(\Phi))/2$. Those are used in *SSAH* for choosing random angle values for each explicit impact. For the implicit cases grouping a large set of impacts, the average angles are used.

### 3.2.2 Spin-up Efficiency: The Angular Momentum Transmitted to the Target

The target has an initial spin vector **ω**, mass *M*, angular inertia tensor **J**, and angular momentum (AM) **h**. The impactor has AM $\boldsymbol{\delta h_i}$ and will create target vector increments of spin $\delta\boldsymbol{\omega}$, a scalar decrement of target mass $\delta M$, increments of angular inertia $\delta\boldsymbol{J}$, and target angular momentum $\delta\boldsymbol{h}$.

The spin-up efficiency $\zeta$ is defined as the ratio of the magnitude of the AM imparted to a target compared to the AM of the impactor. It is a major factor in the history of spins for the asteroids and it has significant uncertainty.  Laboratory experiments are the only cases of actual experimental impact results for the spin-up efficiency.  In those experiments, the targets were small, not spinning, and, depending on the experiment, most of the fragmented mass departed the target.

---

[13] This assumes impacts are equally probable from all directions from the primary Z direction.

So, if the impactor has AM magnitude $|\boldsymbol{\delta h_i}| = RmUSin(\phi)$, the target AM increment $|\boldsymbol{\delta h}|$ is some factor $\zeta(\phi)$ of that and depends on the impact angle $\phi$.  It is written (Burns and Safronov, 1973) using Eq. (18) as

$$|\boldsymbol{\delta h}| = \zeta(\phi)|\boldsymbol{\delta h_i}| = \zeta(\phi)RmUSin(\phi), \tag{22}$$

in terms of that "spin-up efficiency factor" $\zeta(\phi)$.  That factor is generally less than unity because of down-stream escaping mass, although some researchers assume it is unity[14] (the "sticky impactor" assumption).  It might even be greater than unity for elongated objects struck at their extremities.

The few experimental results (Davis and Weidenschilling, 1982; Yanagisawa and Hasegawa, 2000) are sparse and have lots of scatter. They show a definite drop in the value of $\zeta$ as the zenith angle $\phi$ increases from 0° from the vertical (perpendicular to the surface) to 90° (tangential grazing impacts).  For angled impacts into sand, and measuring the horizontal component of linear momentum transferred to a flat surface, Davis and Weidenschilling, 1982 report a near-linear dependence on angle, with the fit

$$\zeta(\phi) = 0.8\left(1 - \frac{2\phi}{\pi}\right). \tag{23}$$

This data was obtained for low impact velocities (1 to 2 km/s) and using low angle impacts into a sample consisting of a horizontal box of sand.

A later study by Yanagisawa and colleagues reported experiments into the flat surface of free-falling cylindrical targets, for several target materials.  For spherical basalt targets they give the formula

$$\zeta(\phi) = 0.41 Cos^2(\phi), \tag{24}$$

but with limited data defining the angle dependence.  These experiments are difficult, and there are many reasons for ambiguity.

There are other factors that likely affect the spin-up efficiency. First, asteroids are not spherical. Better, they might be considered ellipsoidal, either prolate or oblate, with an aspect ratio, the ratio of the longest to shortest principal axis, ranging from unity for a sphere to as much as 2 or 3.  Smaller ones are on average more elongated than smaller ones.

The angular momentum of the impactor, $RmUSin(\phi)$, is based upon a "lever arm" $rSin(\phi)$ from the center of mass to the impact point. For an elongated object that lever arm can be larger (or smaller) than the radius $R$ of an equivalent sphere. Consequently, the values of the transfer efficiency might extend over a wide range. Further, for elongated targets the AM attained by the target can also be enhanced in another way. For a spherical target, the late-stage ejecta cone is perpendicular to the surface, so while it has linear momentum, it does not have any angular momentum. However, for an elongated target that is not true.

---

[14] In principle, if the ejecta patterns are strongly affected by a pre-existing spin (as in the "angular momentum drain" concept), the vector angular momentum of the target need not even be in the same direction as that of the impactor; and it could have a negative magnitude. However, this study did not find any cases where the angular momentum drain was significant. (see § 3.2.3).

Harris, 1990 used a model to study the effects of elongation of a body on a spin-up due to an impact and came to the conclusion that the elongation was of minor importance. Yanagisawa and Hasegawa, 2000 considered that same subject, and came to a very different conclusion. They reported as much as a factor of 4 increased AM transfer efficiency for objects with aspect ratios of 2 to 1. These cannot both be right. And both assumed only infinitesimal increments.

Spin-up efficiency might also depend upon the size (i.e., gravity) of the target asteroid, because of the amount of mass of the target removed from the target in the impact process. Clearly, in an impact, more mass escapes a small target with small gravity than a large target with large gravity. However, whether that has any effect on the target spin is unknown. Harris, 1990 argues that all recoil from ejecta loss can be neglected in terms of altering the spin rate.

Finally, there is an important question about how spin-up efficiency might depend upon impact magnitude. That is another question with little data or calculations to answer one way or the other. Lab experiment of CD events having no initial spin typically produce a remnant core with near zero spin. Historically, (e.g., Harris, 1990, Cellino et al., 1990) assume that a target in a catastrophic event has no more spin-up than a target in a non-CD event.

Taken all together, there are good reasons to believe that the spin up efficiency is greater than given in the above formulas. However, it is probably limited to the value $\zeta = 1$, the so-called 'sticky impactor' assumption. That simple limit is adopted as the default in *SSAH*.

This is an important area needing more study. The overall nature of the simulations of spin up is unlikely to be severely affected, but the modeling of the spin-up efficiency will play an important role in the exact values of the average spin versus diameter.

### *3.2.3. Spinning Targets: "Angular Momentum Drain"*

The details of the removal of mass in an impact event affects the spin imparted by that impact; but conversely, a pre-existing spin may affect the mass removal. In a cratering event, the faster ejecta will escape the target, the slower ejecta will not. The escaping ejecta mass carries away mass and angular momentum.

However, for a spinning target, even a spherical one, the ejection velocity is increased in the direction of the surface velocity (prograde) due to that spin; and is decreased in the other (retrograde) direction. As a result, more ejecta will escape in the prograde direction, and less in the retrograde direction. Irrespective of the impact direction, that difference might decrease or even reverse the spin increment compared to an asteroid with no spin, for both spherical and ellipsoidal objects. That effect for cratering (small) impacts was called the "*Angular Momentum Drain*" (Dobrovolskis and Burns, 1984, hereafter denoted as D&B).

*SSAH* models this process, following the procedure of D&B, but with improvements on the ejecta scaling assumptions. The analysis depends primarily on the values for the total ejecta mass *M\**, and a slowest velocity *v\** ($v_c$ in D&B notation). The previous §3.1.3 defined those variables and the current assumed ejecta scaling.

D&B ignore the cases when only a part of the cone of ejecta of ejecta escapes because of the increased velocity due to the spin[15]. Then, for a single impact, the decrement of spin because of the drain effect is given in their Eq. 25 (their *k* is *3μ* here) as

$$\delta\omega_{drain} = -\frac{15}{12}\mu\omega \left(\frac{m}{M}\right)\left(\frac{R}{r_a}\right)^{-3\mu},\qquad(25)$$

for objects with $R > r_c = \frac{v^*}{\frac{8\pi\rho G}{3}}$, and zero otherwise. Here $\omega$ is the spin, $m$ the impactor mass, $M$ the target mass, $R$ the target radius, $\rho$ its mass density, $G$ the gravitational constant and $\mu$ is the point-source scaling exponent. Their $r_a = r_c \left(\frac{M^*}{m}\right)^{\frac{1}{3\mu}}$ in terms of the ejecta properties.

That formula is included in *SSAH*. However, in running the code, I did not find any case where the drain increment was significant. That result is consistent with results of others (e.g. Farinella et al., 1990). That may be because it is a modification to the magnitudes of spin increments only for cratering impacts, and those themselves are not large contributors.

A strikingly different conclusion was presented by Ševeček et al., 2019 using code calculations for impacts into rotating objects. However, I would argue that their calculation material strength models were not appropriate.

### 3.2.4. Spin Targets: "Splash"

A similar effect of existing spin may occur when a spinning asteroid has an impact that nearly or completely disrupts it. Because of the spin, the pieces that are broken off will do so in a non-symmetric manner, and again that affects the spin increment due to the impact. This process was discussed and modeled by Cellino et al., 1990, and called "*Angular Momentum Splash*". Their model has an initial spin $\omega_i$, a "post impact" spin $\omega_{PI}$ before any splash, and a final spin $\omega_{final}$ after all the fragments are ejected and final re-accumulation has occurred.

Starting with the pre-impact initial spin $\omega_i$, they assume that the first post-impact spin $\omega_{PI}$, before splash, is given from an increment using an efficiency $\zeta(\phi) = \zeta_0 \cos^2(\phi)$ just as for cratering above in (27).[16] They consider a range of $\zeta_0$ from 0.2 to 1. That post-impact spin assumption is separate from their assumptions abut the splash effect, and it is the same as here for the impact spin for large impacts.

For the effects of the fragments, they use a rudimentary model that synthesizes fragment velocity values as a function of distance from a central "irraditation point" at some location, and make assumptions about the decay of material velocity with distance from that point. Using that, and numerous computer analyses of different cases, they obtain ranges of values for how much the splash effect reduces the spin increment from the impact, all in terms of several free parameters that are not well known. To first order, the "splash" spin increment they find is proportional to the post-impact spin and the ratio of the target's decreased mass to its original mass:

---

[15] I have derived a much more complicated expression that did not use that simplification, but I now think that the estimates of this drain process do not require those more detailed results. I did not find any case where the drain is significant. (see also Davis et al., 1989 and Farinella et al., 1990).
[16] They denote this as $\beta(\theta)$.

$$\delta\omega_{splash} \sim \omega_{PI} \frac{\delta M}{M}. \qquad (26)$$

where the multiplying coefficient is of order of unity, although they discuss other possibilities. The increment of mass is negative, so the increment of spin is the opposite sign as the initial increment. Basically, they assume that this increment is late-stage and independent of the original direction and mechanics of impact.

Here their approach is followed. The spin $\omega_{PI}$ is calculated just as for cratering impacts. That calculation uses the original angular intertia to connect AM to spin. Then the increment (28) is used to get final spin. (That equation includes both the change of AM and inertia.).

However, just as for Momentum Drain, I did not find any case where the splash increment was important. In any case, the code *SSAH* has an input knob that adjusts the spin-related models for testing their effects in more detail.

### *3.3. Initial Values*

A simulation of the history of the main belt asteroids requires initial values for the diameters and spins. In many cases, a simulation is for a defined collection: a well-defined group such as a family, or the set of objects in a spin database. However, an existing database has the present diameters, not the initial ones. If a simulation begins with a set of 'current' diameters then, after any defined time, they will be smaller, substantially so for the smaller ones. To account for this difference, the code can be first run with these database diameters as initial diameters, and then a new starting population is found by beginning with diameters increased by the amount they decrease during that trial run. An iteration can be used if required. *SSAH* has preprogramed initial diameter sets for the final diameters of the objects in the 2019 spin data base.

The initial spins must also be defined. The importance of those depends upon the asteroid size. All small asteroids are destroyed on a timescale much shorter than the history of the main belt; and, because of their replacement in the population, their spin distributions approach a steady state on a relatively short time scale (see §6.3.3). For that reason, their initial spins are not important, although the assumed spin for the replacement does matter some (next section). And the very largest, above 100 km, do not have their spin changed appreciably during its Solar System history, assuming the population has been steady in time. For that reason, their present spin is likely a primordial spin. While that initial spin could be put in the simulations as an initial spin, it would just carry through to the final spin distribution. But there may be some effect for intermediate-sized objects. In *SSAH* the initial spins are arbitrarily set to random small values between $\pm 10^{-6}$ and $\pm 10^{-5}$ radians/s (0.015 to 0.14/day) and very small wobble.

### *3.4. Reset Values*

The analyses here are based upon the assumption that the population of the main belt is steady in time. To maintain that property, when an object is pulverized it is replaced by another one using the initial diameter of the lost object. That the new object is presumably a fragment from a parent body that was one or two orders of magnitude larger.

That new object's spin must also be assigned. It should be a random choice from a defined distribution of the fragment spins. However, the knowledge of the spin of fragments is sparse. Giblin et al., 1998 used buried explosive charges as a proxy for impacts on strength-dominated, 21 cm diameter, alumina cement objects, with zero spin. The fragments averaged about 2 cm diameter. At that size scale, all fragments have very high spins.

A theory for the scaling of those spins would be required to predict spins for a large, and for perhaps a gravity-controlled body. That is another area of uncertainty in the simulations. I resort to an empirical approach.

The cross-plot of spin magnitudes in a limited diameter range was shown above in **Fig. 4**. The average spins are not affected much by the spins assumed for a depleted objects replacement, but the slow spins in this distribution do. Therefore, they provide guidance for the spin reset model. In that plot, the diameter range is from 5 to 10 km; so all spins are below the gravity spin limit at 10/day. There is a broad distribution with a peak at about 3 to 4/day. Then there are a large number of very slow spinners.

In *SSAH* I set half of the reset spins randomly below 0.1/day, and the other half to a random value between 0.1 and 3/day. Recall those are only their starting spins, they are then spun up during their lifetime. I will show the final simulation histograms below in §6.3.3.

## 3.5. From Angular Momentum to Spin

There is another important issue about a spin analysis. In the transformation from the angular momentum to the spin, the angular inertia of the body is required. However, the mass and inertia are not constant through an impact process; they are changing concurrently with the changing spin over a very short time increment. There are three different plausible inertia values to use. First is the initial inertia of the object before the impact. That would seem to be the appropriate choice if the entire body had a change of spin on a shorter timescale than the mass loss and CD processes.

Second is the inertia including both the original target and the mass of the impactor, although for any surviving asteroid, the mass of the impactor is very small compared to that of the target. Some previous papers (e.g. Davis et al., 1979; Harris, 1979b; Dobrovolskis and Burns, 1984) included the added mass of the impactor. The contribution in slowdown due to the increased inertia from the impactor mass was called the "mass-loading drag". Here I ignore that.

Finally, there is the inertia of the object after all the ejecta mass and fragments have escaped.

*SSAH* uses the original angular inertia of a target.

## 3.6. Spin Limits

There are physical limits on how fast an asteroid can spin. Those exist because spins create internal stresses, which have limited magnitudes. For small asteroids, There are spin limits depending upon the material strength of the object. For large asteroids, the self-gravitational stresses are much larger than material strength values, and those gravity stresses limit the spin of an asteroid. Thus, as is also the case for other aspects of solar system impact physics, spin limits have two distinct regimes. Small asteroid spins may be limited by a strength limit, and larger asteroid spins are limited by a gravity limit.

### 3.6.1. Gravity Limits

Perhaps the most striking feature of the distribution of the spins of asteroids is the occurrence of a limit spin at about 10 cycles per day (about 7 $10^{-4}$ rad/s, 2.4 hrs./cycle), for all asteroids with a diameter of 1 km or greater. Hartmann and Larson, 1967; Burns, 1975; Harris and Burns, 1979; Harris, 1996 and others observed that that value is the rotational breakup limit for aggregates with no tensile strength (now called "rubble piles").

It is very important to recognize that in a geological material the absence of tensile strength does not mean the absence of shear strength. It is not uncommon to confuse those two strengths. The difference is dramatically illustrated by the obvious difference between the behavior of dry sand and the behavior of water. Holsapple, 2001, 2004 derived an analytical theory for the spin limit for ellipsoidal solids using the Drucker-Prager model of soil mechanics. That model has zero tensile strength, but not zero shear strength. It has the pressure-induced shear strength of a granular material. I solved analytically for the dependence of the spin limit on the shape and bulk mass density of the body.

An example result showing the dependence of the limit on asteroid shape and properties is presented here:

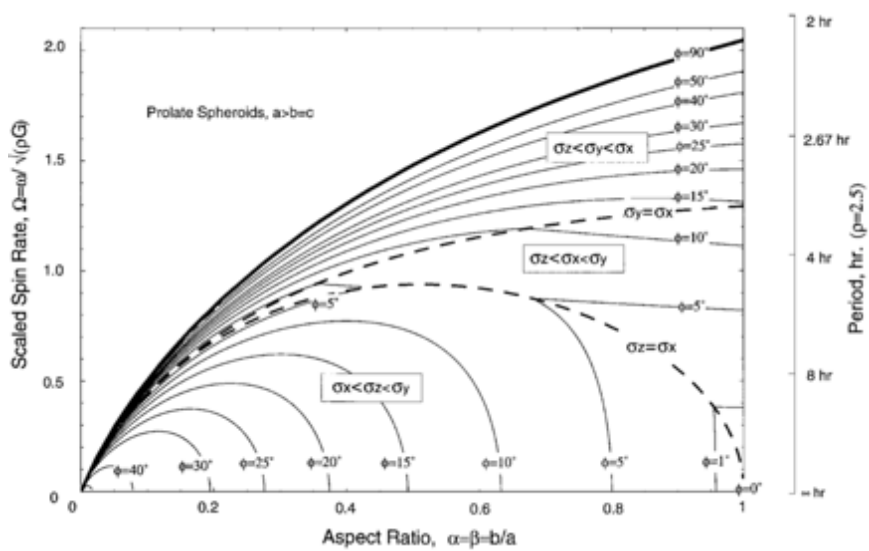

**Figure 12.** Analytical spin and period limits as a function of the aspect ratio of a prolate object. For a mass density of 2.5 g/cm³, soil-like angles of friction, and near-spherical objects, the range of limit periods is from about 2.4 to 3+ hours (8 to10 cycles/day). Less ideal shapes can be expected to have smaller limit periods. The database has an upper period limit of about 2.3 hours. (Holsapple, 2001).

Many authors have used the observation of that spin limit to infer that all larger asteroids are rubble piles, with no tensile strength. As an example, a 2000 publication stated: "The cutoff, …, are evidence that (asteroids) larger than a few hundred meters are rubble piles". The limit is often called "the cohesionless spin limit". Many similar statements exist in the literature.

Those statements are not correct. Holsapple, 2007 extended the limit theory to consider spin limits for objects with tensile strength. He showed that the above limits restrict the spin of all larger asteroids irrespective of their strength; because the strength expected for any sufficiently large asteroid is smaller than its self-gravity forces. Said differently, that limit is a gravity limit, it always dominates strength limits for large objects, and applies to all large asteroids regardless of their strength. Zero tensile strength is not required nor indicated. Simply, it does not matter. Unfortunately, many researchers still state the incorrect interpretation.

There is also a question about the history of the objects that do exceed the spin limit. They cannot remain with a spin above the limit; but must readjust their configuration and return to, or below, the spin limit. How that is done will depend in some way on its properties. Perhaps, for small spin increments not much above the spin limit, surface regolith might slide towards the equator and even be cast into space, forming an oblate, "top-shaped" object, such as the asteroids Steins, Ryugu, Bennu and others. The angular inertia would increase, and the spin would slow. Whether that is possible would depend upon the initial shape and the value of the shear strength in the regolith surface.

Alternatively, for a large spin excursion above the limit, more major changes might occur. One might imagine a tendency of a highly elongated body to split into two or more objects and form a binary object, or perhaps an asteroid cluster. That would depend upon global measures of strength in the interior of the target. We do not yet know the details of those processes.

In the *SSAH* simulations, the spin must be reassigned after a gravity limit excursion. Since there is not enough known about the actual asteroid shapes and their response to excess spin, those are set lower in a random way.

### *3.6.2. Strength spin limits*

The gravity spin limit applies to the spin of the larger asteroids: the cases where the gravity stresses dominate the material strength of an object. However, the gravity stresses are negligible for the smaller asteroids, and material strength values could limit the spins. Those limits would be above the gravity limit. Beginning about 25 years ago, increasing numbers of such cases have been found. The small asteroids spinning above the gravity limit are called the "fast-rotating-asteroids", or "small-fast-spinners".

As mentioned above, Holsapple, 2007 derived spin limits based upon material strength. Those limits are included in the **Fig. 3** and other plots of spin distributions here. In Holsapple, 2007, average and maximum locations of the small fast-spinners along a power-law line similar to, but below the spin limits in a plot of spin versus size were taken as evidence that the spins were limited by a strength. However, the simulations here (§ 6.3.4) will show that is not the correct interpretation. The strength limits do not constrain the small fast spins. The maximum spin magnitudes that can be generated are less than those material spin limits for other physical reasons. Therefore, the spin data cannot be used to infer those small asteroids have small strengths (see §6.3.4).

### *3.6.3. The Transition Range*

The spin data verifies that the gravity spin limit applies for all asteroids greater than about a km diameter; and, apparently, few or none below about 200 m diameter. That fact requires that the gravity stresses dominate any governing strength for the larger ones. But asteroids are not homogeneous: they have variable internal strengths and different strengths in any existing regolith. For such reasons, there is apparently also a transition region, from about 200 m to 1 km, in which some asteroids have strength dominating over gravity, and others do not.

In addition, there are questions about the nature and quality of the actual data in that transition region. As noted in §2.3, there is data in that region, gathered using wide-field methods; but its accuracy has been questioned. I have chosen to sidestep questions about the data quality in that region and include only the few reported to be of high quality.

That arbitrary choice also requires an arbitrary choice about how to model that transition region. There is nothing in the theory presented here that defines in a meaningful way the data in that region, except that it clearly lies between the fully gravity and fully strength regions. The code adopts a direct way of restricting the number in a simulation of objects in that region. Only a designated (small) percent of objects set by an input choice are allowed to have spins above the gravity limit in that diameter range. The choice of that percentage will determine the number of objects with spins above the gravity limit in that transition size range, and also the sharpness of the spin distribution curve between the strength and gravity regions. All in all, that percentage choice is some proxy for unknown spin readjustment mechanisms and strength values, although the connection is nebulous.

## 4. THREE-DIMENSIONAL SPINS: WOBBLING

The simulations construct not only spin magnitudes, but a complete three-dimensional spin vector history. The spin increments will have components in all three directions relative to the principal-axes (PA) coordinate system of any asteroid. All asteroids will have some amount of wobbling. Therefore, define the number of robbers one must define the criterion used. In the in the intervals between impacts, the wobbling will decay toward zero due to internal dissipation. The important questions are about the amount of wobbling, how it is changed in an impact, and the timescale for which it decays to a near-zero amount.

### *4.1. Wobble Decay*

The wobble angle of a target as a function of time is denoted by $\beta(t)$. In terms of the components of the spin $\{\omega_x, \omega_y, \omega_z\}$, it is given as

$$\beta = ArcTan\left(\frac{\sqrt{(\omega_x^2+\omega_y^2)}}{|\omega_z|}\right). \tag{27}$$

Over a time-interval of interest, it has an initial value $\beta_0$ and, of its own accord, will decrease (damp) in time with some timescale $T_0$. Apparently the first to study the wobble nature and the time-scale for a return to a Principal Axis (PA) spin was Prendergast, 1958, with a number of studies since that time[17].

The standard scenario has an object attaining a tumbling state from an impact and then, over some characteristic timescale $T_0$, relaxing towards a PA rotation. However, in fact it is not that simple: over that relaxation time many more impacts are likely to occur. That is a dominant feature in the simulations reported here. And each impact changes the spin and the wobble. In the time interval between each impact, the decay to a principal axis spin is asymptotic in time. The history of every asteroid is one of recurring wobble states and multiple decays towards a PA state. For a simulation, we need the history of wobble beginning at any initial value $\beta_0$. That will define the spin history until the next impact, at which time there is a new $\beta_0$ and a new decay.

---

[17] Sharma et al., 2005 give a reference list.

The theory is not entirely developed for application to the history of each individual case. The focus has been on the timescale $T_0$, and not on the actual history of wobbling. Generally, that timescale is assumed to depend on a "quality factor" measure $Q$: an energy loss per cycle due to dissipative effects. $Q$ is most often taken as a material constant, independent of the frequency or magnitude of the tumbling. Using that, several groups have derived results for the damping time. Sharma et al., 2005 discusses those previous analyses.

### 4.2. The Wobble Timescale $T_0$

The relaxation time $T_0$ for any wobble history is defined here as the time for the wobble to decay from some initial value $\beta_0$ to some definite final angle $\beta_f$. Assuming the time scale does not depend on the spin or the size, and on the product $\mu Q$, then a simple dimensional analysis[18] shows that the timescale has the form:

$$T_0 = A(shape, \beta_0, \beta_f) \frac{\mu Q}{\rho D^2 \omega_0^3}, \qquad (28)$$

where the variables $\{\rho, D, \omega_0, \mu, Q\}$ are, in order, the mass density, the target diameter, the rotation rate, an elastic modulus[19], and the quality factor. The function $A(shape, \beta_0, \beta_f)$ is a function of the geometry (shape), and initial and final wobble angles $\beta_0$ and $\beta_f$.

The value of the final wobble angle is chosen as some assumed smallest detectable wobble[20]. Henych and Pravec, 2013 state that tumbling is photometrically detectable from light curves when the rotational axis misalignment angle is greater than about 15°. Others (e.g. Efroimsky and Lazarian, 2000) used a threshold of 6° for detecting wobbling, as found in radar observations. *SSAH* allows the user an input choice, but it only is used to define the number of wobblers.

### 4.3. Wobble History

To first order, the wobble decay rate depends linearly upon the instantaneous wobble value. Then the history is a decaying exponential, which can be written as

$$\beta(t) = \beta_0 e^{-Ln(\beta_0/\beta_f)(t/T_0)}. \qquad (29)$$

The simplicity is apparent. For this exponential function, for given initial and final wobble, the entire history of the wobble is determined by only one variable, the relaxation time $T_0$. The results from any study can be used to define that relaxation time; but there are significant variations in the choices of initial and final wobble angles. Also, all are based on a choice for the elastic modulus and quality factor, which are not well known.

---

[18] The dimensional analysis assumes that there is a dependence on the product $\mu Q$, and not on each term separately.
[19] Note that historical precedents lead here to a dual use of the symbol $\mu$, and of the symbol $Q$. Above here $\mu$ was the PS scaling exponent, here it is an elastic modulus. Previously, $Q$ was the impact energy per unit mass, now it is the quality factor.
[20] The value for $\beta_f$ is only used to define $T_0$. *SSAH* actually calculates the wobble down to 1°, which is then reset to 0°.

The simple numerical expression given by Harris, 1994 with converted units is:

$$(T_o)_{yr} = \frac{1.1 \, 10^{-3}}{\omega^3 D_{km}^2}, \tag{30}$$

with $\omega$ is in radians/sec. There is no dependence on the initial or final wobble in that expression, which cannot be true for the specific definition of decay time used here. I assume it is for an initial wobble of 45° and a final of 6°. The expression given in Sharma et al., 2005 for an initial angle of 45° and a final of 6° is about twice that value. The default in *SSAH* is taken as (30) for those choices of initial and final angle. Then the value of the decay time for any other initial and final angles $\beta_0$ and $\beta_f$ is

$$(T_o)_{yr} = \frac{1}{2} \ln\left(\frac{\beta_0}{\beta_f}\right) \frac{1.1 \, 10^{-3}}{\omega^3 D_{km}^2} \tag{31}$$

Therefore, a decay from 45° to an angle of 1° takes 1.9 times longer than from 45° to an angle of 6°. That default value of the final wobble angle that defines the relaxation time can be adjusted in *SSAH* by the user in the interface pane. Its value and the decay model will determine the number of wobblers, which are put on the output plots and can be compared to the data. This analysis gives a new way to estimate wobble decay times from actual data, using specific relaxation times dependent on both the initial and final wobble angles.

### *4.4. Free Spin Decay*

In the intervals between impacts, an asteroid is freely spinning in space, with a decaying wobble. The conservation of angular momentum dictates that the spin must change as the wobble angle changes. Suppose an object begins with the spin vector $\boldsymbol{\omega} = \{\omega_x, \omega_y, \omega_z\}$ where $z$ is the direction of the primary principal axis. The spin will evolve over some time interval $\Delta t$. Over that time interval, the lateral components $\{\omega_x \quad \omega_y\}$ and the $\omega_z$ component are assumed to change to values $\{\gamma_1 \omega_x \quad \gamma_1 \omega_y\}$ and $\gamma_2 \omega_z$. A decay time is defined from (31) and the new wobble from (27), in terms of $\gamma_1$ and $\gamma_2$. Those two values are then found by solving for the new wobble assuming the constancy of angular momentum. The solutions are found using elementary algebra, although they are quite lengthy. The code calculates those values.

That completes the analysis of the change in spin during the interval between two impacts. Then the new spin is obtained by adding the impact increment to that existing spin.

When the time interval between impacts is greater than the decay time, the spin increasingly relaxes to the PA spin (the wobble is cutoff below 1° in the code), and the final spin components become

$$\omega_{final} = \text{Sgn}(\omega_z) \left\{ 0 \quad 0 \quad \frac{\sqrt{(I_z^2 \omega_z^2 + I_x^2 \omega_x^2 + I_y^2 \omega_y^2)}}{I_z} \right\}. \tag{32}$$

as given by Burns and Safronov, 1973. For a prolate body with c/a=0.6, the numerical result is

$$\omega_{final} = \text{Sgn}(\omega_z) \left\{ 0 \quad 0 \quad \sqrt{(\omega_z^2 + 0.36(\omega_x^2 + \omega_y^2))} \right\}. \tag{33}$$

From this, the net spin and the lateral components can be seen to always decrease, while the PA component always increases.

# 5. THE *SSAH* IMPLEMENTATION

The code *SSAH* generates recreations of main belt asteroid histories using the models presented above. Each model is coded in a separate function, thus providing an easy structure for model improvements. A detailed outline of the *SSAH* code is given in Appendix A. Here a summary of the approach is given.

The input panel (shown in Appendix A) lists all the input variables, and each one has a button labeled "Huh?" that explains it's meaning. There is a set of pre-defined input files that can be used, either as-is or as the start to modify for new runs. Each new input set can be saved in a file for future use.

*SSAH* runs with a set of (target) asteroids defined by the user. That set can be defined in various ways in the input panel. They are drawn from the assumed size frequency distribution and determined at the outset. Each target has a pre-defined initial size, spin, type, gravity limit, spin-up efficiency, and nominal aspect ratio; all chosen in a stochastic way. From those their mass, angular inertia, and other properties are calculated. The objects of that set are tracked in time through multiple impacts

The calculations are divided into time steps. The population is held constant over the time step, and it can be updated at the end of each time step. Thus, the time stepping is a form of a forward-difference Euler method. Movie frames as requested are generated at each timestep. If one assumes a steady impactor population (as for the results presented here) then the time steps are only important for the movies.

In each time step, there is a primary Do-loop over the designated target asteroids. For each target asteroid, the expected number of impactors from each of the binned population is calculated. The number of expected small impacts can be enormous. For that reason, the expected impactors are divided into two categories: the large 'significant' ones and the remaining smaller ones. The division is based upon the energy of the impactor per unit target mass $Q$ compared to CD threshold $Q^*$ of the target, a typical value is $Q > 10^{-3}Q^*$. An explicit calculation is done for each of those significant impacts. Impacts with smaller magnitude, $10^{-3}Q^* > Q > 10^{-6}Q^*$ are calculated using an implicit form with averaging all in each bin. The even smaller impacts are ignored.

The above Table 1 can be consulted for the values of those significant impactors. For example, for a 1 km C-Type target, all 1200 impactors with diameter greater than 4 m are deemed significant and calculated individually in detail. 5000 smaller ones are included in the implicit group. For a 100 m target, there are 5000 significant impacts, and tens of thousands in the implicit set.

At each time step and for each target, *SSAH* must determine the diameter range for the explicit impactors and for the implicit impactors. The intrinsic collision probability for the time step defines the number of expected hits of each set from those diameter ranges. For the explicit impactors, that expected number is drawn randomly from the population and its defined diameter range. Each of those explicit impactors is assigned random values for defining characteristics: its velocity, density, impact time within the current time step, and impact angles. That set of explicit impactors and properties is sorted by time of impact and recorded in an array (list).

Following is a second loop over that set of explicit impactors. In their time order, the explicit effects of each impact are calculated. Spins exceeding the gravity limit are randomly reduced to below the target's spin limit (likely cases of spin-off of mass or binary formation). Wobbling is relaxed over time according to theory on wobble decay (§4.2). Then the results using the bin average of all the implicit impactors, times the total number of implicit impactors, is added.

The compute time for a *SSAH* simulation[21] depends substantially on the target population chosen. Simulations with more small objects require a lot longer run time but add nothing to results for the larger objects. One useful simulation uses every fourth object from the 2013 spin data, which has 1677 objects. The simulation on my 3.6 GHz 8-Core iMac is completed in about 10 minutes and uses about 05. GB of memory. The entire set of the ~17,000 asteroids in the 2019 spin database is simulated for 4.0 Byr in about 3.5 hours and again takes about 0.5 GB of memory

Further information about the code is given in Appendix A. Further, at the beginning of *SSAH* there are buttons that link to files and plots with further information.

## *5.1. Material Constants*

The models have various material property constants. The asteroids have been divided into two types, C-types, and the S-types. Those two classes are expected to have different values for the material properties. While many of these properties are not well known, informed choices have been made for the values used in the *SSAH* code. Using the notation introduced above, the Table 2 below lists the values currently in the code. There are too many of these values to include inputs in the general input panel, but they can be edited in the function 'asttypestuff' portion in the source code. The calculations are done in kg, km and sec units although some time units are input as Byr.

| Constant | C-Type | S-Type | Notes |
|---|---|---|---|
| $\mu$ | 0.41 | 0.55 | Scaling exponent of the point-source measure |
| *dens* | $1.5 \times 10^{12}$ | $2.5 \times 10^{12}$ | Mass density, (kg/km$^3$) |
| $Y_0$ | $1.5 \times 10^9$ | $7 \times 10^9$ | Lab-sized strength in (10), (kg/km/sec^2) |
| $d_{0strength}$ | $10^{-4}$ | $10^{-4}$ | Lab specimen diameter $D_0$ in (10),(km) |
| *nsize* | 4 | 4 | Size-Strength coefficient *n* in (10) |
| $nsize_{spall}$ | 2 | 2 | Spall Strength varies with $D^{1/nsizespall}$ |
| $K_{vS}$ | 0.3 | 0.3 | Ejecta scaling constant in (13a) |
| $K_{vg}$ | 0.3 | 0.3 | Ejecta scaling constant in (13b) |

---

[21] A skilled Mathematica programmer could undoubtedly speed up the calculation considerably. I have made no attempt to do so.

| $K_1$ | 0.15 | 0.05 | Strength regime cratering coefficient in (9) |
| $K_2$ | 1. | 1. | Gravity regime cratering coefficient in (8) |
| $k_R$ | 1.2 | 1.1 | Crater Radius constant in (11) |
| $qc1$ | $2\ 10^{-3}$ | $1\ 10^{-3}$ | Strength regime $Q^*$ definition in (3) |
| $qc2$ | 1. | 1. | Gravity regime $Q^*$ definition in (3) |

**Table 2.** Material constants for cratering, dispersion, and ejecta models in *SSAH*.

## 6. SIMULATION RESULTS

*SSAH* calculates detailed histories for any prescribed set of asteroids. It exposes many features of asteroid histories. The primary results are the for erosion and of spin.

### *6.1. Erosion: Diameter and Mass Histories*

The present large-velocity impacts in the main belt are always erosional, so the size and mass of each asteroid always decrease in time. Because of the stochastic nature of the simulations, each simulation and each history are different. Moreover, they are vastly different depending on initial asteroid size. Further, while the general form is well determined, the time scale does depend on the chosen population.

**Figs. 13a to 13f** present typical mass erosional histories grouped for six initial sizes of targets. These are not meant to evaluate averages, which can be quite different in each simulation; they are presented to show the variety of histories. The green star symbols indicate a CD, red markers at the ends indicate a pulverization and termination of that object. Since the histories are all independent, the results for a grouping of 10 at a time is the same as the results calculating each of those separately. The variety of histories is a result of the stochastic methods of the simulations.

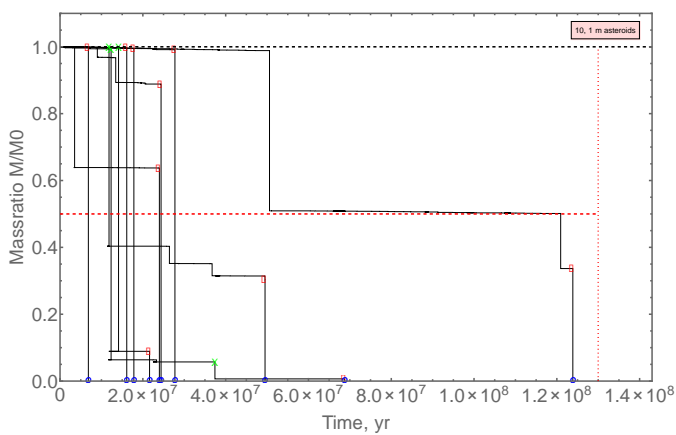

**Figure 13a.** Mass (size) histories for ten initially <u>1 m asteroids</u>. In every case a few large impacts entirely dominate the erosion, the numerous small impacts are almost imperceptible at this scale (There are a large number showing if one magnifies this plot). All ten are gone by 130 Myr.

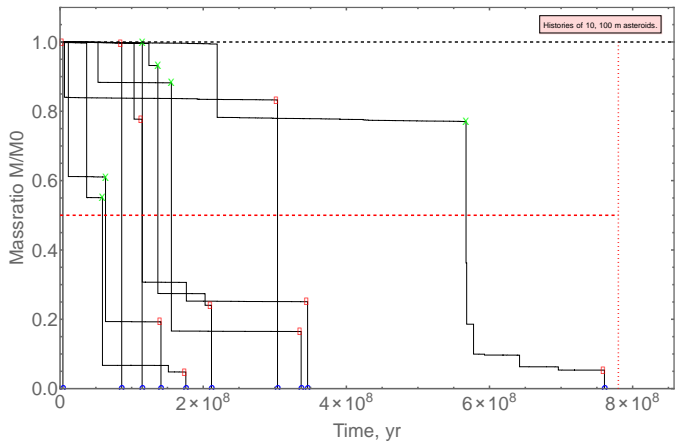

**Figure 13b.** Mass (size) histories for ten initially 100 m asteroids. In each case, a few large impacts dominate the erosion, the numerous small impacts are almost imperceptible. Most are gone by 300 Myr, one lasts 800 Myr. One is reduced to less than half-mass by erosion before being dispersed.

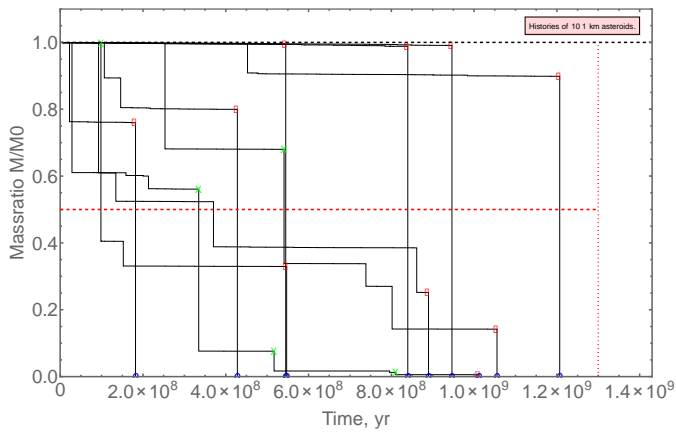

**Figure 13c.** Mass histories for a set of ten initially 1 km asteroids. They last up to 1.3 Byr. The few large impacts are dominant, mass loss from small impacts is again imperceptible at this scale, and not important to the erosion.

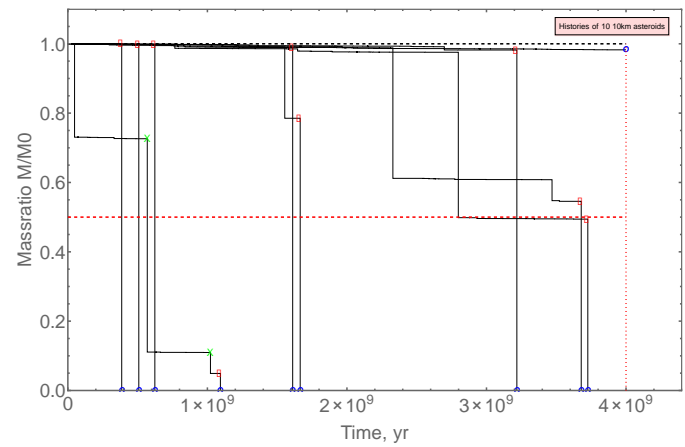

Figure **13d.** Mass histories for a set of ten initially 10 km asteroids. Only one survives 4 Byr. The few very large impacts are dominant, small impacts are not discernable. One is reduced to half mass by erosion alone.

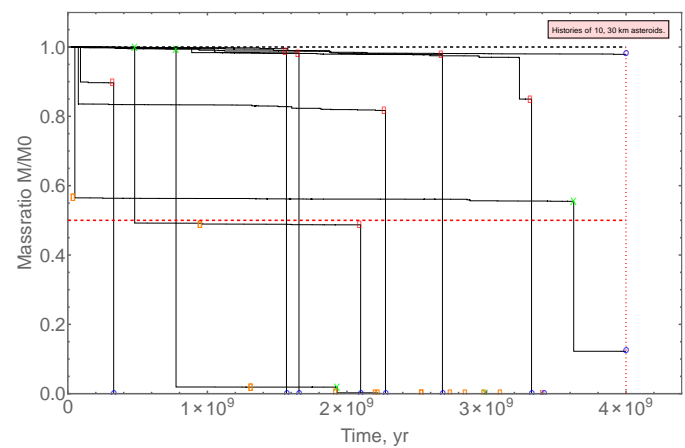

**Figure 13e.** Mass histories for ten initially 30 km asteroids. Only 2 survive 4 Byr. Only one avoids a major impact.

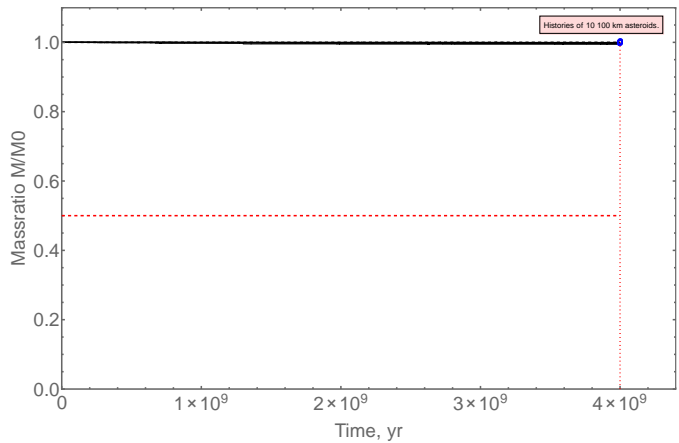

**Figure 13f.** Mass histories for a set of 10 initially 100 kilometer asteroids. There are no CD or pulverizations, no noticeable mass loss and all survive the 4 billion years.

All these results show the same characteristic features:

• Significant size changes result exclusively from a few large, near-catastrophic impacts: total erosion from all small impacts is negligible.

• All asteroids with diameter less than a few kilometers survive for only a small part of the history of the main belt. All except the largest of the current asteroids are the result of fragmentation of larger objects. Thus, their current size or spin history distributions provide no information about the earlier population of impactors.

• Asteroids with a diameter greater than 100 kilometers all remain relatively unscathed even for the 4.5 Byr of the belt, assuming that the present population was present. A small percent may be disrupted to form families (see §7.3). Presumably, these large asteroids are a product of accumulation in the accretion phase of the solar system.

### 6.2. Individual Spin Histories

The spin histories also provide additional important insights. **Figs. 14a-14f** depict spin magnitude histories for sets of 10 targets with six different initial diameters. The green star symbols indicate a CD, red markers at the ends indicate a pulverization. The sign indicates the ± orientation of an assumed principal axis component.

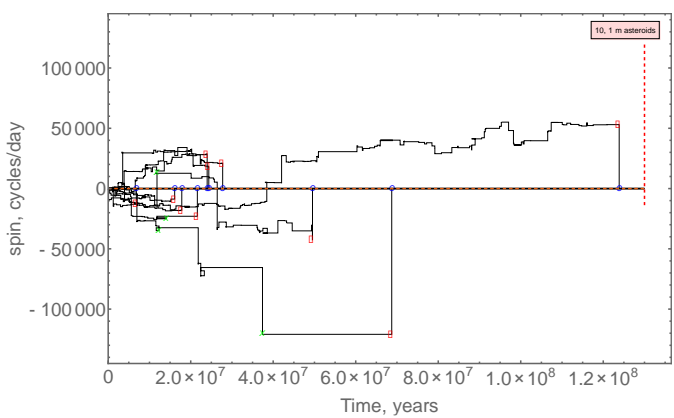

**Figure 14a.** Spin magnitude histories for 10 small asteroids with initial diameter of 1 m. These are all small, fast-spinners. Note the extremely large values for the spins: as fast as 1/second. The initial spin-ups occur in a few Myr, a small part of their lifetimes. Each history ends with a pulverization before 130 Myr. On average spin magnitudes increase with time, but there are a few spin reductions and reversals.

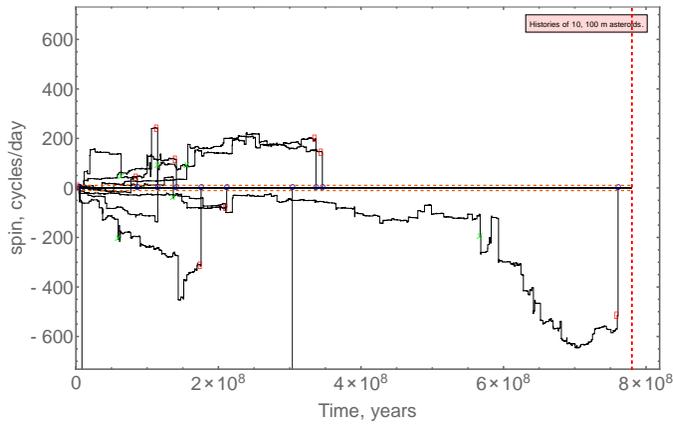

**Figure 14b.** Spin magnitude histories for 10 asteroids with initial diameter of 100 m. These are all small-fast-spinners, here up to 40/hr. Each history ends with a pulverization before 800 Myr. On average spin magnitudes increase with time, but there are a few spin reductions and reversals.

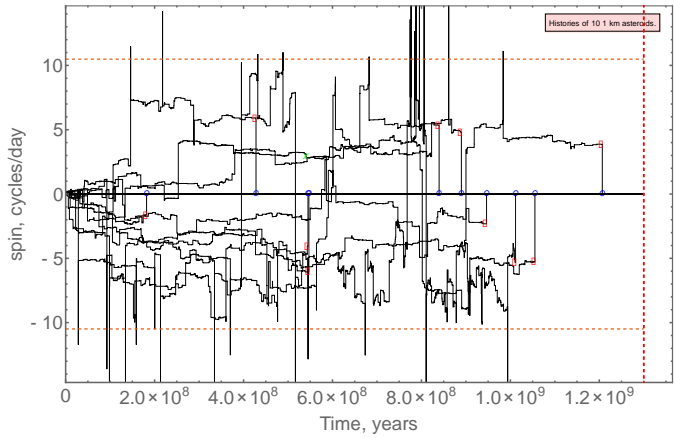

**Figure 14c**. Spin histories for 10 asteroids with initial diameters of 1 km. The spins are limited to the gravity spin-limit indicated by the red dashed limits at about ±10 cycles per day. The vertical "spikes" and instant spin reduction are excursions above that gravity limit. In those cases, the target must undergo some mass reduction or reshaping to increase its angular momentum and reduce the spin to below that gravity limit. Note multiple gravity limit excursions and occasional spin direction reversals.

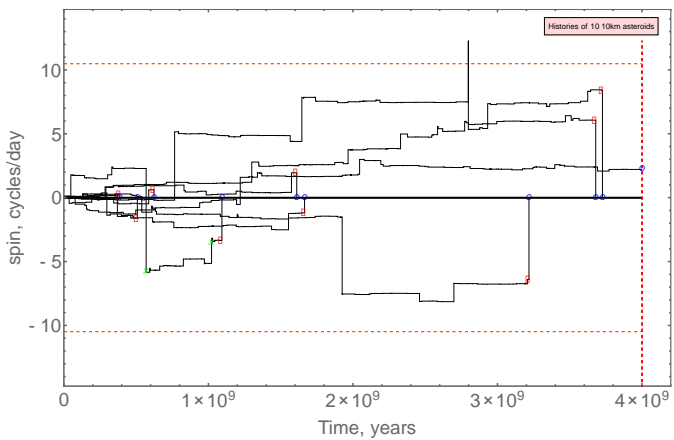

**Figure 14d**. Sample spin histories for 10 asteroids with initial diameters of 10 km. Only two exceed the gravity spin limit, after they first have been reduced in size.

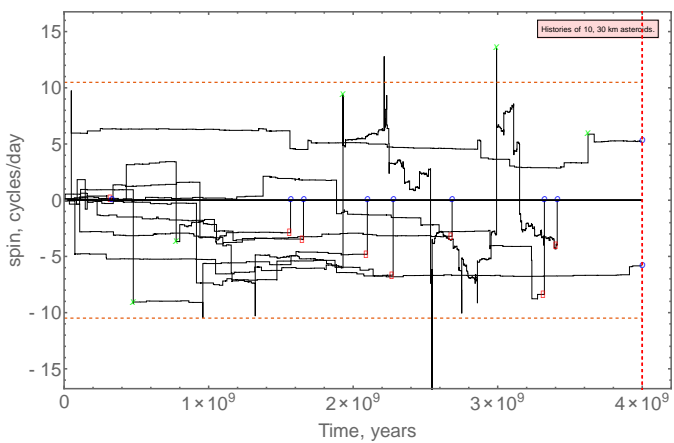

**Figure 14e**. Sample spin histories for 10 asteroids with initial diameters of 30 km. Three exceed the gravity spin limit. The spin of large asteroids is limited by a lack of significant impactors, not much by the gravity spin limit. There are spin reversals. All significant spin increments are a result of a large impact; the large number of small impacts accumulate small spin increments.

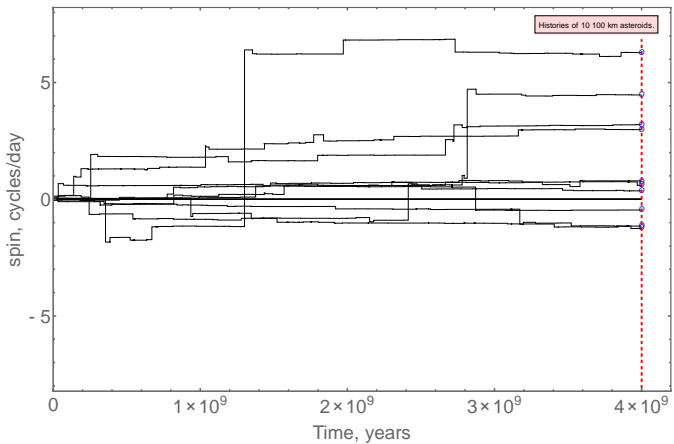

**Figure 14f.** Sample spin histories for 10 asteroids with initial diameters of 100 km. Several achieve spins of several/day. All significant spin increments are from large impacts.

These simulations depict a history from an initial very small spin to a pulverization. In many other types of simulations, when an object becomes pulverized it is replaced in the population at its original size but with small spin, and this spin-up and breakup process is repeated all over again. They net effect of this cycling of the spins is to evolve into an equilibrium process, see §6.3.4.

For all small targets, the nature of that repeating process is not affected by the population of impactors, but its time scale is. That is because the rate of spin-up and the rate of destruction are both changed by the same factor when the population is increased or decreased. Any target large enough to avoid destruction over most of the life of the solar system escapes that cycle.

General observations include:
- Small asteroids are spun up to periods as small as seconds.
- Spins are not limited by any physical material strength (discussed in §6.3.5).
- The spin increments are dominated by a few large impacts, the spin accumulation from the multitude of small impacts is small. Spin theories based on a "random walk" quadratic accumulation of infinitesimal increments are not applicable.
- Small (<1 km) asteroids are typically spun up in a few tens of Myr, a small part of their total lifetime. For that reason, few small asteroids are spinning slowly.
- Until becoming close to the gravity limit, most often the spins increase in time, especially from the larger impacts.
- It is the asteroids with diameters from 1 to 10+ km that spin-up to and over the gravity limit. These are the candidates for the formation of tops and binary objects (see §6.3.5).
- Asteroids may exceed the spin limit several times.
- This study considers spin increments from impacts only; the results conflict with assumptions that YORP spin-up is required to produce spin-up and binary fission.
- The fall-off in maximum spins for the large asteroids is due to an impactor supply deficit, it is not due to any physical process constraints.

## *6.3. The Spin Distributions*

The major test of *SSAH* simulation fidelity comes from comparing the outcomes of a simulation using a target set with the same diameters as the asteroids in the actual spin database[22]. The impactor population used was discussed in §2.1. The 2019 spin database version used here (Warner et al., 2019) has 17,115 objects in the main belt and NEO space. The diameter range is from 4 m to the 945 km Ceres. The data for asteroids larger than about a km are mostly from the main belt, while that for the smaller ones are mostly from the NEO space. Comments supporting combining those data are presented in §8.

The input panels for a simulation are shown in Appendix A. There were 40 time-steps over 4. Byr. A steady population, KAH-SFD was assumed, so after each pulverization a target was replaced at its initial size, with the spins explained in §3.4. Pulverizations happened 41 $10^3$ times total, multiple times for each small target. There were $14\times10^6$ impacts calculated explicitly, an average of 846 per target. There were 28,000 CD dispersion events, and 16,000 reductions to less than half-mass by CD, erosion, or pulverization. 20.6 % experienced gravity spin limit occurrences, those are candidates for binaries, multiples or 'top-shaped' configurations. At the end, 3361 or 20% were tumblers with at least 15˚ rotation tilt from the principal axis. The entire calculation took about 5 hrs and used about 700 MB memory[23].

### 6.3.1. The Spin Distribution.

**Fig. 15a** are the data for the spins versus diameters of the 2019 database (repeated from **Fig. 3**). **Fig. 15b** is the result of the simulation from *SSAH*.

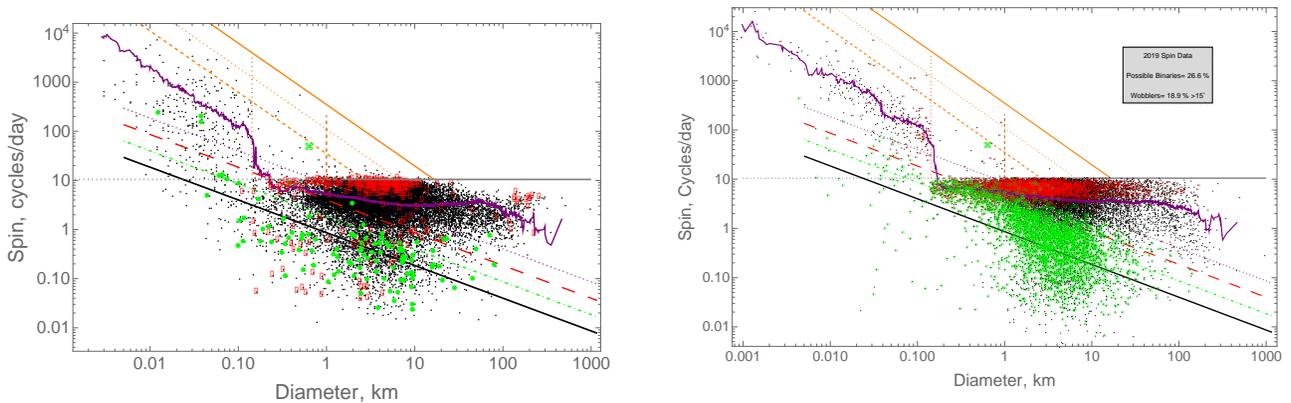

**Figures 15 a, b**. The final spins versus the diameter after 4 Byr of the ~17,000 objects of the 2019 spin database. The spin data is on the left, the simulation result is on the right. The jagged purple lines are running box means of the data over a ±10% range of diameters[24]. For the actual data, the 320 red symbols are binaries, while for the simulation the red symbols are targets that have experienced a spin exceeding the gravity limit and are candidates for binaries or tops. This simulation has 26.6% such points, many more than the 2% identified binaries in the data. The green markers indicate tumblers, in the simulation there are 19%, many more than the 4.6% identified in the data.

The overall character of the simulation matches the data very well, an indication that the models are generally correct. It reproduces the two major features of the spin data: the gravity-limited spin of all asteroids larger than 200 m diameter, and the increasingly faster spins for the asteroids smaller than 100 m diameter, the "small, fast spinners". The first feature is the obvious result of imposing the gravity spin limit. The 'fast-spin' feature is due to an absence of any governing spin limits and a result of the paths that are followed in this spin-diameter plot; in which a vertical increase in spin is generally accompanied by a horizontal shift to the left in diameter (see § 6.3.5 below).

---
[22] With slightly increased initial diameters to offset erosion, as described above.
[23] Using 1 core on a 3.6 GHz Intel Core i9 iMac.
[24] In contrast to some previous studies, all spins were included.

The spins of the largest objects at the right, above a few tens of kilometers, generally fall increasingly below the spin limit. That is caused by a lack of impactors in the population sufficiently large to greatly affect their spin. (see **Fig. 14f**, and further discussion below).

The wobblers, defined here as those with wobble more than 15°, are indicated by the green symbols. They can be compared to the four lines sloping down from the left to the right across these plots. Those lines are the loci of conditions for which the tumbling recovery times, from the top down: are 4 million, 40 million, 400 million, and 4 billion years respectively. There are some slow spinners among the larger objects, most of those are wobblers. The data has the asteroid Mathilde in that region.

There are only a dozen objects with spins above the gravity limit in the transition range of 0.1 to 1 km diameter; the data has the asteroid 2001 OE84 and a few others. But recall that the number included in the database in this region was arbitrarily restricted to a few, according to quality assessments. Correspondingly, the simulation artificially restricted the number they could be spun up above the gravity limit in that diameter range.

At the left, all small (<100 m) asteroids spin fast; the smallest, with diameters of a few meters, at almost 10 per minute. And, importantly, none of the small fast spinners have a spin which is reasonably limited by a material strength (the upper orange line, see §6.3.4). There are only a few small asteroids with small spins, in the lower left region of these plots. While one might imagine that is because of the difficulty of observation of such objects, the simulation suggests they do not exist. A small asteroid that begins with near-zero spin is rapidly spun-up to a rate of at least several a day and it continues to spin up to a period of minutes or less until it is destroy by a large impact. For most of their lifetimes, they are spinning quite fast. They do not break apart from excessive spin before being dispersed by a large impact.

There are distinct differences in this comparison between the numbers of possible binaries and the number of wobblers. The binary objects are discussed further below in §6.3.6 and the tumblers in §6.3.7.

### 6.3.2. Mean Spins v. Size

The comparison of the running box mean spin versus diameter between the data and the simulations provides a more detailed measure of the fidelity of the reproduction. It is more affected by the parameters chosen for the models. There are three aspects of the modeling which bear on the magnitude of those means.

The magnitudes are determined from an equilibrium between increasing spins in time versus time to a CD event. A uniform change in the numbers of objects in the SFD will change the time scale to that equilibrium, but will not change the values. However, a detailed change in the shape of the SFD does have an effect.

I began this study using the Bottke et al. 2020 population estimate. However, the resulting spins universally had a dip in their mean just below 10 km that is not in the actual data. Testing divulged that the dip was a direct result of a dip in the SFD for objects around and less than 100 m diameter. The SFD for sub-km asteroids is not well determined. Therefore, a modified SFD designated as 'KAH-SFD' was constructed to bridge that dip, changing the values from the Bottke et al. 2020 SFD for the extrapolation below 1 km. It was presented above (**Fig. 1**). That was successful in removing the dip in the spin results, but the overall values were still lower than the data, by factors of 2 to 3.

The remaining two models that directly affect the mean spins are the spin up efficiency model, and the CD Q* curve. Either might be considered a free parameter, but within limits. Various combinations can likely be constructed to match the mean spin curve with the data. However, there are reasonable constraints which make some changes seem more likely than others. In the results presented here, my choice, which gives quite good results, is to assume the simple 'sticky impactor' assumption, in which the entirety of the impactor angular momentum is transferred to the target. The $Q^*$ curve was retained as presented above.

More detailed changes would improve the comparison, for example, a falloff in spin up efficiency above 20 km would likely remove the hump at 50 km. (Perhaps because those targets are more spherical on average compared to the smaller ones?). I do not pursue those details here. Until such time is there a more definitive data about either of these models, it cannot be said which combination of changes is to be preferred. And recall that the database is a very incomplete snapshot of the entire family of asteroids in the belt.

**Fig. 16** compares the means of the spins for the data and for the simulations:

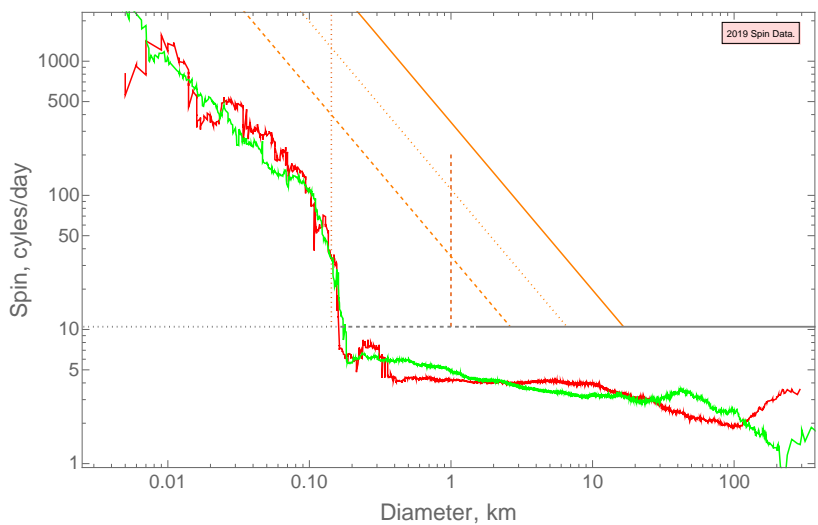

**Figure 16.** A comparison of the final running box means of simulations (green curve) compared to the data (red curve). The general character is matched well, however there are some differences, especially for the smallest (<10m) and for the largest (>100 km) objects.

The simulations of the fast spins of the asteroids smaller than about 10 m diameter are always above those of the data. Most likely, that is a selection effect of the spin database. The simulations suggest there are many more very fast spinning objects with diameters smaller than 10 m and spinning at rates of several per minute and higher to be found. Note that the number of small objects in the spin database are but a small fraction of the total number of asteroids in that size range, so there are many more to be found.

And at the upper end, the data has a dip centered at 100 km, and then an upturn, but the simulation spins are always decreasing above that diameter. That "V-shape" feature in the data has been discussed in the literature. It has been believed since the 1950's, that the "V-shape", and the minimum at ~100 km, is real.

This observation has been interpreted in several ways. Kuiper et al., 1958; Anders, 1965; Hartmann and Hartmann, 1968; Tedesco and Zappala, 1980 all suggested that the smaller bodies arise through collisional fragmentation, while the larger, faster-spinning objects are primordial. Farinella et al., 1981 suggest that ellipsoidal asteroid shapes could account in part for slower spins. Dobrovolskis and Burns, 1984 propose the angular momentum drain process and Cellino et al., 1990 and Farinella et al., 1981 suggest that the splash effect may account for that feature, although their numerical simulations did not support their proposals.

Here the earlier proposals are mostly supported. The sub-100 km asteroids may or may not be fragments, but they are spun up to their present values over the lifetime of the main belt. The spin-up falls off dramatically for those larger than 100 km. The 'V-shape' minimum in the data is the intersection of a fall-off in spins occurring from the present population with original larger spins created from either a primordial state, or from a prior history for which the population had more large objects.

### 6.3.3. Spins in a diameter range

A final and most detailed comparison of a simulation with the actual spin data compares more than just the average, but also the distribution around the average for a given range of diameters. The above **Fig 4** presented that data from the database, for the diameter range of 5-10 km. Here I reproduce that plot and compared to that same measure for a simulation.

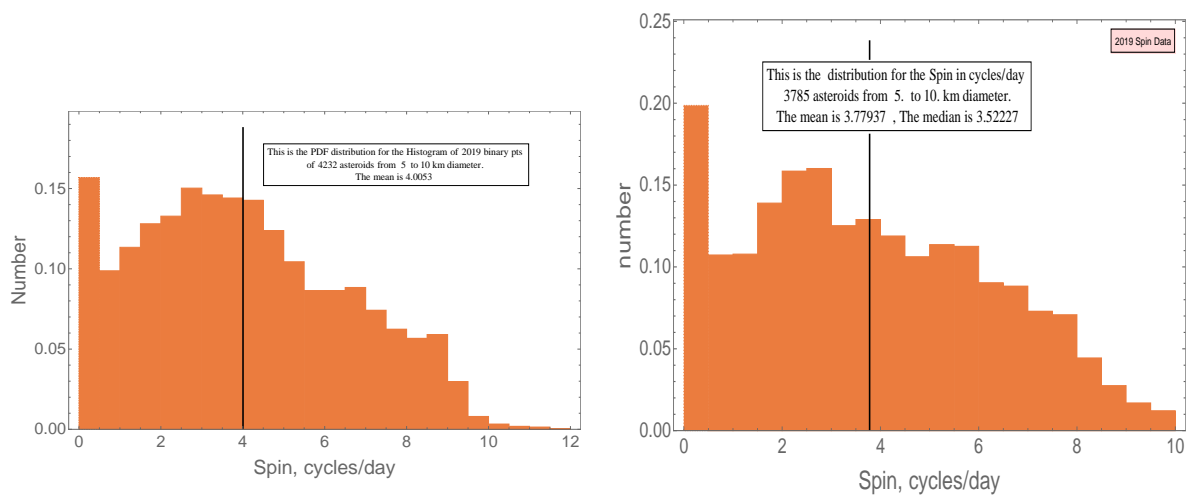

**Figure 17.** On the left is a distribution of spins around the mean of the 2019 spin database. On the right is that from a simulation of that same data, stretched vertically to align the ordinate values.

This outcome is affected by the choice of the spin values to assign to objects being replaced in the steady population after a CD. The average of this distribution is not much affected by that choice, but the distinctive nature of the large number of very slowly spinning objects is affected. Since average spins increase in time, the slow spinners can only be for fragments of very recently shattered objects. Therefore, the spins of fragmented objects must include a significant number of very slow objects. SSAH assigns small spins to 50% of them.

### 6.3.4. The Time Evolution of the Spin Distribution

The simulations here assume that the current population has been steady for some length of time. But how long might that be? It cannot be for more than the approximate 4.5 Byr lifetime of the main belt after its formation. But might it be for only a few hundred Myr? Researchers typically suggest periods of several Byr.

The results above compare a stimulation for 4+ Byr to the current spin database. But what would that comparison look like if the duration of the steady population was significantly less? Does it match equally well for much shorter peroids? What are the various time scales that produce the current spin distribution?

To address those questions, one can compare spin distributions for at various periods of time. Those are obtainable from the movies produced by *SSAH* showing that distribution versus time. I've selected snapshots of that movie and reproduce important ones. **Figs. 17** are those snapshots.

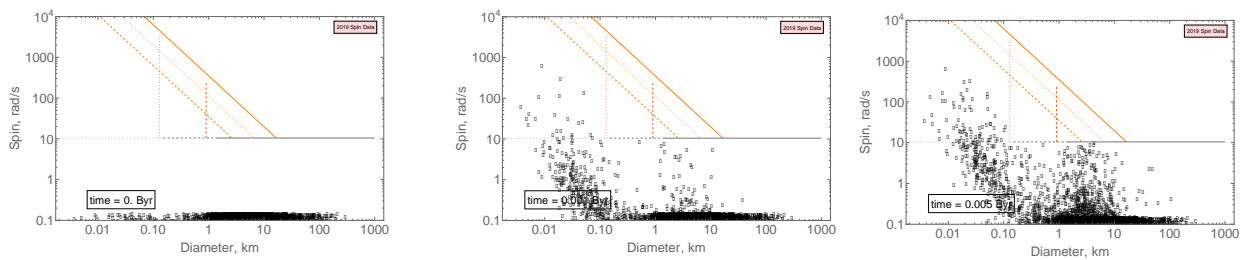

**Figure 17a.** This first row of frames is for the early times: the first 5 Myr. The first frame indicates the initial state, with all objects assumed to have very small spin. By the second frame, at 1 Myr, many objects have achieved substantial spin, especially the small fast spinners. The third frame, at 50 Myr has the first few km-sized objects approaching the gravity limit, more fast spinners and increasingly fewer small slowly spinning asteroids.

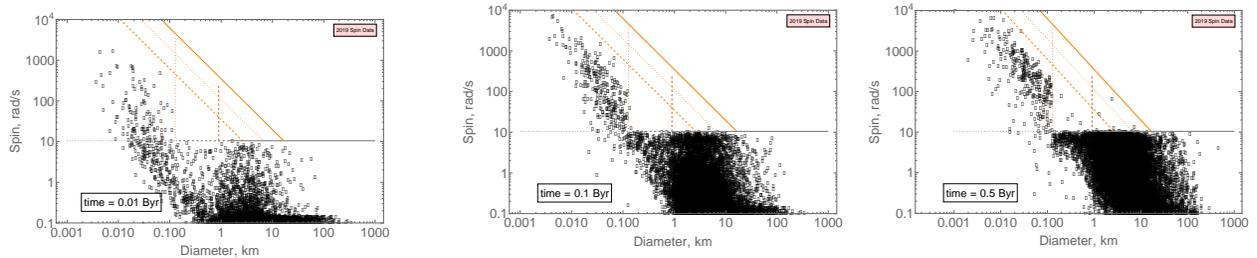

**Figure 17b**. The evolution of the spin distribution from 10 Myr to 500 Myr. By the time of 10 Myr, several km-sized asteroids have reached the gravity limit., but none over 10 km diameter. The small fast spinners are grouped more tightly, with even less slowly spinning objects.

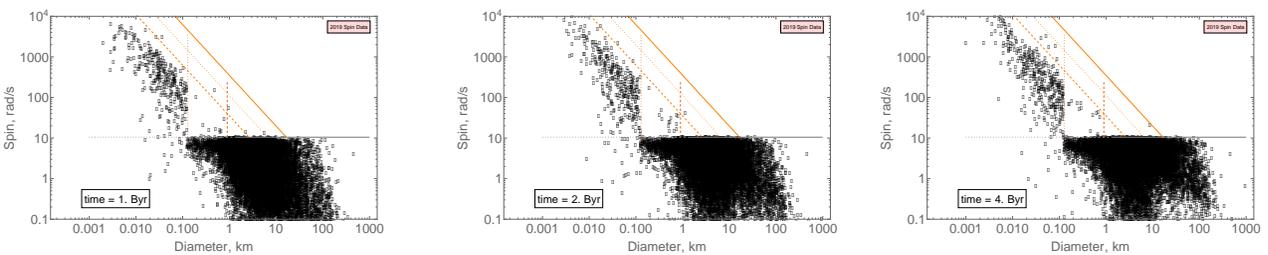

**Figure 17c**. The evolution of the spin distribution from 1 Byr to 4 Byr. Since 100 Myr there are few observable differences. The distribution has come to equilibrium.

The spin distribution comes to equilibrium relatively quickly, in a few hundred million years. Therefore, in that data there is no obvious information about the duration of a steady population. However, there are perhaps detailed changes. Not shown here is the fact that the number of excursions above the spin limit, which might lead to binaries, increases continuously over these entire times.

### *6.3.5. The Role of the Spin Limits*

The spin of an asteroid creates internal stresses; if those stresses exceed the permissible stresses, a mass readjustment must occur. (see §3.6)

As discussed above, the spins of all larger asteroids are limited by the gravity limit, but those of the smaller ones are not. However, are those smaller ones limited in spin by a strength limit discussed in §3.6.2? The importance of spin limits can be discovered by comparing a simulation with the limits to one without.

First, simulations were made for the 2019 spin data objects with the gravity spin limits suppressed. Without the limit, the spins increase to as much as 30 cycles/day, well above the approximately 10/day gravity limit value. The gravity limit definitely limits the spins.

Then a simulation for the 2019 spin data objects was rerun with the strength limits inactive. The results were indistinguishable from the runs with strength limits in **Fig. 15b**. There are only a few objects above even the 1% strength line, and no spins anywhere near the expected 100 % strength limit line.

Therefore, for the asteroids below about 1 km diameter, where the magnitude of a strength limit has been derived, the objects do not reach those strength limits, even using unrealistically low values of the strength. For those asteroids, the calculated spins with no limits are the same as those with strength spin limits. Simply put, they are not spun up to values allowed by a strength limit. Why not?

Consideration of the details of the history of the spin versus the diameter reveal the reasons. For the larger objects, an impact produces only a small mass change, so the spin increases while the diameter remains essentially constant. In a diameter versus spin plot, the history paths are essentially vertical, until the spin is limited by the gravity limit or the object is dispersed.

However, for the smaller objects, for example 100-meter objects, the histories are very different. **Fig. 18** shows the path followed by 40 initially 100 m asteroids in a plot of spin versus diameter, compared to the strength spin limits derived in Holsapple, 2007.

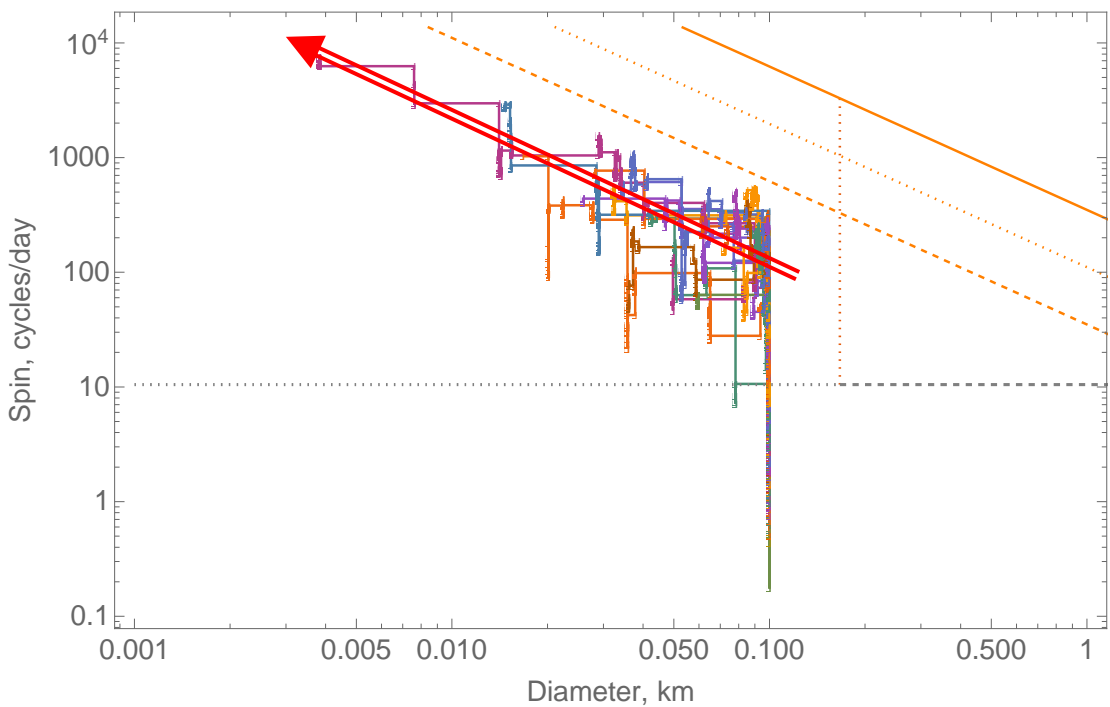

**Figure 18**. The

evolution of spin versus diameter of a set of 50 initially 100 m asteroids ('small fast spinners'). The average paths followed for the larger increments are approximated by the bold red arrow, which avoids the strength limit lines (orange lines). Each path ends when that asteroid is pulverized. The slope of the red arrow in this log-log plot is about -5/4, the same as the strength limit lines. No spin reaches a strength limit.

None of the spin-diameter paths come close to any of those strength limits, even the limit assuming only 1% of the expected strength. The reason is apparent. For all significant spin increments, the resulting impacts impart not only additional spin but also considerable mass loss and diminished diameters[25].

In Holsapple, 2007, the average and maximum locations of the small fast spinners along a power-law line was provided as evidence that the spins were limited by a strength spin limit. But it was also noted the actual spin values were all more than two decades below the curve using the expected limit for typical geological materials. So, in fact, at that time there were really two possible explanations: either the strength was much less than expected, or the data was not constrained by strength at all. Holsapple, 2007 overlooked the second explanation and incorrectly endorsed the first explanation. Since then, several research groups have presented arguments and models about why asteroids could have very small strength. The spin data gives no information about the strength of small asteroids except that it must be greater than the gravity stresses.

### 6.3.6. Numbers of Binaries and Tops

As has been stated several times, asteroids larger than about 1 km in diameter cannot sustain spin faster than the gravity limit. An object spun up past that limit must undergo a morphological change to increase it angular inertia and slow its spin. Some might become binaries or top-shaped. But there are significant differences in the number of objects of the simulations with spins exceeding the gravity limit compared to those currently identified as binaries in the data.

Binaries have only been recognized in the last three decades. Weidenschilling et al., 1989 questioned whether they even existed. By 2015 they were thought to be relatively common (Margot et al., 2015), and formed by spin-up and fission processes. It is stated in the literature (e.g. Pravec and Harris, 2007, Margot et al., 2015) that the NEA population contains about 15% binaries, and that it is likely that there is that same percentage in the main belt. However, the 2019 database shown in **Fig. 3** currently only identifies less than 2% of the total as probable binaries. In contrast, the simulation shown above resulted in 25-30% or more of the objects having experienced spin exceeding the gravity limit.

The simulations impose the gravity limit for all asteroids larger than about 1 km in diameter. The code retains the information for all asteroids exceeding the spin limit, and simply assumes that some change must have occurred to slow the spin to a value below that limit. While the nature of that change is, as has been stated, uncertain; those that exceed the limit are candidates for formation of binaries, or of asteroid clusters, or formation of 'top-shaped' objects. It may be that only part of the gravity limit excursions produce binaries. Or perhaps the simulations are running longer than a steady population duration?

---

[25] The interested reader might derive the ratio of the spin increments to the diameter increments for these small asteroids for a significant impact.

The binaries among the largest diameter asteroids are most likely a result of a large impact and re-accumulation, not a spin up.

Clearly this is an area needing more study, but the simulations provide useful insights.

### *6.3.7. Wobbling and Tumbling*

The spin distributions of the data given above in **Figs. 15** and the output from the simulations identified about 19% asteroids wobbling more than 15°. As was discussed in § 4, all asteroids may be expected to be wobbling by some amount. The simulations use a default value of 15° to identify a wobbling state, but that is adjustable by the user in the input; and further, even in the output plots. Recall that a value of 6° is popular in the literature, that would increase the numbers of designated wobblers (see **Fig. 21** below). The wobble relaxation time also has a significant effect, in *SSAH* it also can be changed from the default value given in Eq. (30).

To compare the simulation to the data, there are a couple of issues to be considered. The first is the uncertainty in the wobble value that can be (or has been) observable. It may be difficult to assess wobbling and spinning motions from the light curves. Furthermore, there is an observational selection aspect; wobbling objects are a significant fraction of smaller asteroids with a long rotation period – both properties unfavorable for photometry (Pravec et al. 2014).

In these simulations, the entire history of tumbling of each target is calculated using the theory given in § 4. In **Fig. 19**, I present one plot of the wobble history for 700 Myr of a single small, initially 5 km asteroid. It was destroyed and replaced several times in its solar system history, each time it was reborn with zero wobble. Wobbles are updated for each impact and then until the next impact decay at a rate depending on the objects size and spin, as defined in Eq. 30. Since the time scale of new impacts may be on the same order as the relaxation time, wobble increments are often added before a return to near PA spin. An occasional large impact produces a large wobble and can also change the sign of the spin. During epochs of large spin, recovery to near PA spin is almost instantaneous on this time scale.

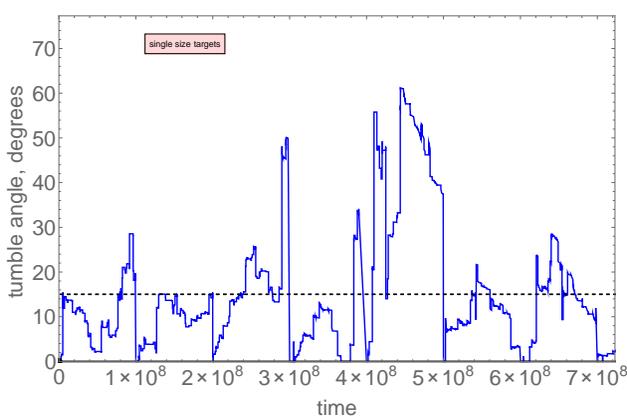

**Figure 19.** The wobble angle history of a 5 km asteroid over 700 Myr. The sudden vertical increments are due to impacts, the downward sloping portions are a decay. During most of this time, this asteroid is spinning slowly (less than 1/day), so its wobble typically decays in about 50-100 Myr, according to Eq. (30). However, when its spin increases, its decay to near zero wobble is much faster. This asteroid has a substantial wobble over much of its life.

In the simulations there are always more wobblers than the 3.5% identified in the spin data, although the exact threshold value for wobble identification in the data is uncertain. If those data are complete, then the wobble decay time may be shorter than the value provided in Harris, 1994. Simulation tests show that there is roughly a halving of that percent for every decade reduction in relaxation time. Alternately, there may be many more wobblers to be identified in the data. The simulations generate interesting clues, illustrating another area for further research.

# 7. ADDITIONAL APPLICATIONS

The simulations can also be used to provide important new insights into other secondary features; including lifetimes, crater counts and family formation. These additional features deserve expanded exposition of their own, but that is not feasible here. I give a short synopsis of each one. At the same time, I present some important new material and interpretations.

## *7.1. Cratering*

In *SSAH* a simulation can be made with a single target. In that case, individual craters can be tracked, using the crater scaling models described above. The craters are segregated into three types. Historically, both bowl-shape (smaller) and complex craters (large) have been recognized. Later (Holsapple and Housen, 2013) the occurrence of spall-dominated craters (small, brittle targets) was identified. Complex craters of moderate or larger diameter do not form on the asteroids. For small targets, the spall craters are an important part of the craters.

*SSAH* calculates all these according to the theory and identifies the type of each crater formed. The spall craters occur only in small asteroids, brittle materials such as the S-types, and only for craters larger than any regolith depth. Their shape is very different than a bowl- shape (Holsapple and Housen, 2021). For example, the **Figs. 20** and **21** illustrate spall crater formed in rocks.

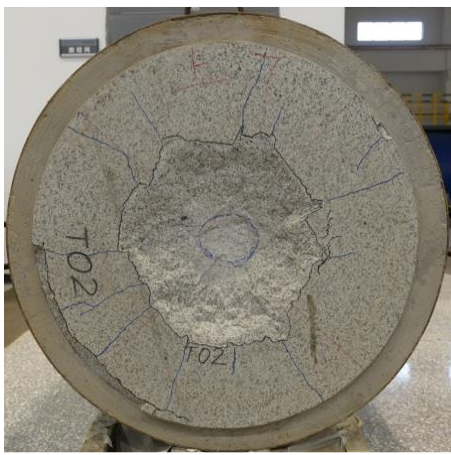

**Figure 20.** Results of a 3.55 km/s impact into a granite target in a laboratory. On the surface of Gaspra, all craters up to 500 m diameter are projected to have this shape: a small central excavation crater surrounded by a large diameter surface spall region.

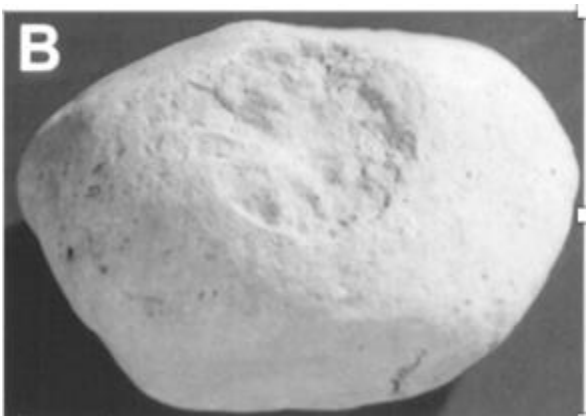

**Figure 21.** A terrestrial laboratory simulation of a 10 cm diameter crater created by the impact of a 1/8" glass sphere at 6.0 km s$^{-1}$ into a limestone boulder. (Hörz et al., 1999).

Compared to a bowl-shaped crater, spall craters are substantially less deep and have about 3 times the diameter. That simple ratio can be used for the size scaling of spall craters. On small S-Type asteroids such as Gaspra, all except the few largest (greater than a few hundred m) are predicted to be spall craters. If true, that will have a large effect on the estimated surface cratering lifetime. This application of the spall-crater theory to an asteroid is new. The reported observations in their literature did not consider nor account for spall cratering.

Unfortunately, other than small craters on earth, there is little actual evidence to support the idea that the small craters on brittle asteroids are spall craters. Some of the evidence would include very broad shallow craters, fewer small craters then expected using standard bowl-shaped scaling, and a large number of surface boulders as a result of the near-field launching of spall fragments.

The **Fig. 22** identifies the approximate range of predicted diameters for spall craters as a function of their surface gravity.

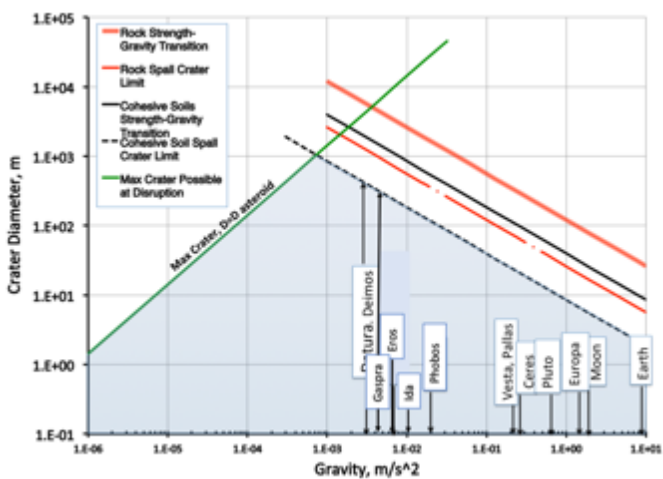

**Figure 22**. The morphology of a crater is based on its size and on the surface gravity of an asteroid. The shaded region shows the predicted range of spall crater diameters as a function of the asteroid gravity. On small S-type asteroids, for example 12 km Gaspra, all craters smaller than a few hundred meters are predicted to be spall craters. Those are a subset of the strength regime of cratering. On the larger asteroids, only relatively small (compared to the diameter) are predicted to be bowl-shaped. On Earth, only m-scaled craters.

These limits can also be plotted as the ratio of the crater diameter to the target diameter, as a function of the target diameter. A plot of that type is below in **Fig 23.**

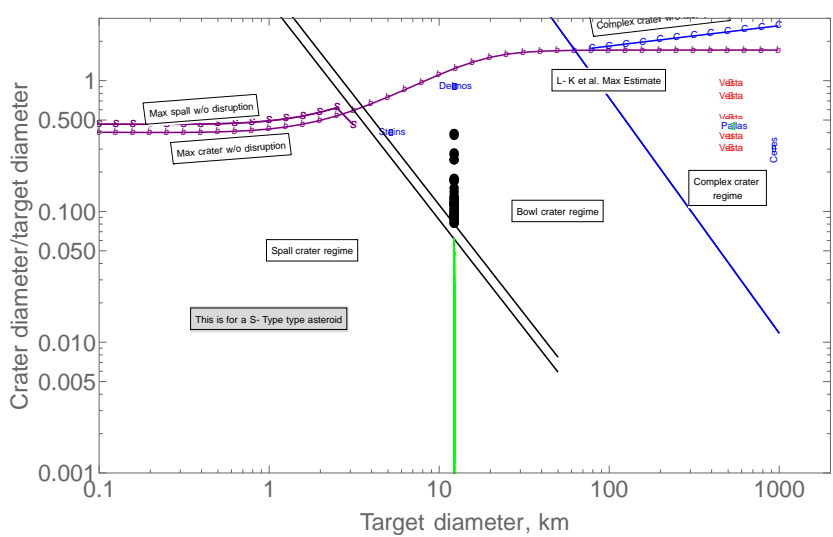

**Figure 23**. The crater to target diameter ratio ranges for each type of crater. At the top (purple and blue curves) are the largest craters that can be formed without catastrophically dispersing the target. The downward sloping lines designate the boundaries between the spall, bowl and complex cratering morphology types.

The values for the largest possible craters without dispersing an object, for each of the three morphology types, are new here; and based on the detailed modeling considerations given above. The broad brown line is the limit given by Leliwa-Kopystyński et al., 2008. The green markers in the spall region, and the black markers in the bowl region at 12 km diameter are from a simulation here for the Gaspra asteroid (see below).

## 7.2. Asteroid Craters: Gaspra

To study cratering on an asteroid, *SSAH* can be run with one single target with a tracking of each individual crater. The crater size distribution can be determined[26]. One such case is the 12 km diameter, S-type asteroid Gaspra (**Fig. 24**), for which there is good cratering data (Thomas et al., 1994; Chapman et al., 1996).

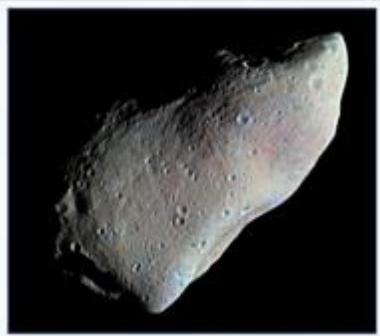

**Figure 24.** The S-Type Gaspra asteroid, a 12 km object with a rotation period of 7/hr. The largest craters are on the order of a few km, but there are possible impact features called 'facets', that are about 10 km diameter.

Referring back to **Fig. 23**, the vertical line of green and black markers at a diameter of 12 km are craters from the results for a simulation of Gaspra. The green markers are spall craters while the black markers are bowl shaped craters. There are about 20 craters larger than one km in diameter, up to a largest of about 6 km. In the discussions of the surface of Gaspra, Chapman et al., 1996 note that there are large 'facets' on the surface, but state that they are much larger than the largest crater possible. According to the scaling here, and **Fig. 23**, that statement is not correct. The upper curve in **Fig. 23** allows a crater as big as the 12 km diameter of Gaspra without catastrophically dispersing it. From to the present scaling, the large facets could be due to impact processes.

For a cratering calculation, *SSAH* outputs include a movie of cumulative crater counts at the end of each time step, so the history of that count is exposed. A comparison of those counts with the Gaspra data provides a means to estimate the crater retention age. The cumulative crater counts from a simulation at 600 Byr are in **Fig. 25**, compared to the data[27] reported by Chapman et al., 1996. This indicates that the surface cratering age for Gaspra is about 600 Myr.

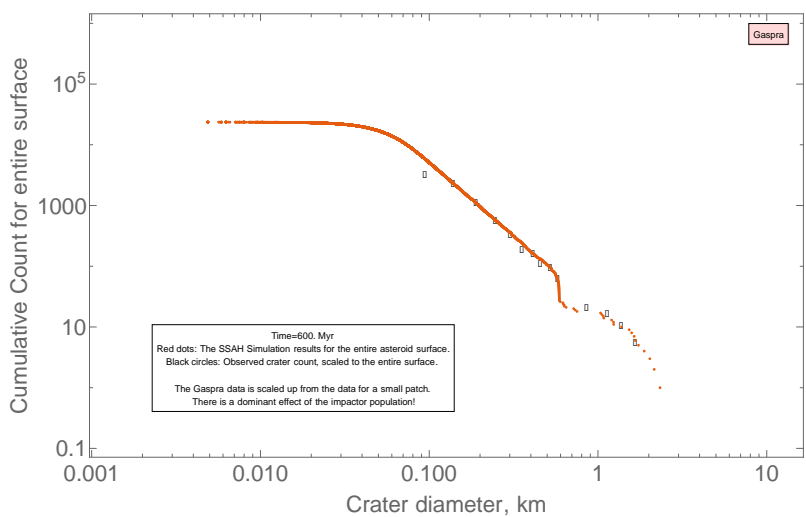

Figure **25.** Crater counts at 600 Myr for the asteroid Gaspra on its 467 km² surface. The red dots are from a *SSAH* simulation, the open circles are from observation data on a 90 km² patch scaled up to the entire surface area (Chapman et al., 1996). The significant "bump" in the simulation just below 1 km diameter is due to the transition from bowl shapes for the larger craters and spall craters for the smaller ones. The match is very interesting.

---

[26] For a single object, the *SSAH* simulation runs very quickly Therefore, to obtain more accurate crater distributions to small crater sizes, *SSAH* increases the range of explicit impact calculations.

[27] The counts of that data are multiplied by a factor of 5.2 to scale from the patch of 90 km² use for the observations to the entire approximately 500 km² of the Gaspra surface.

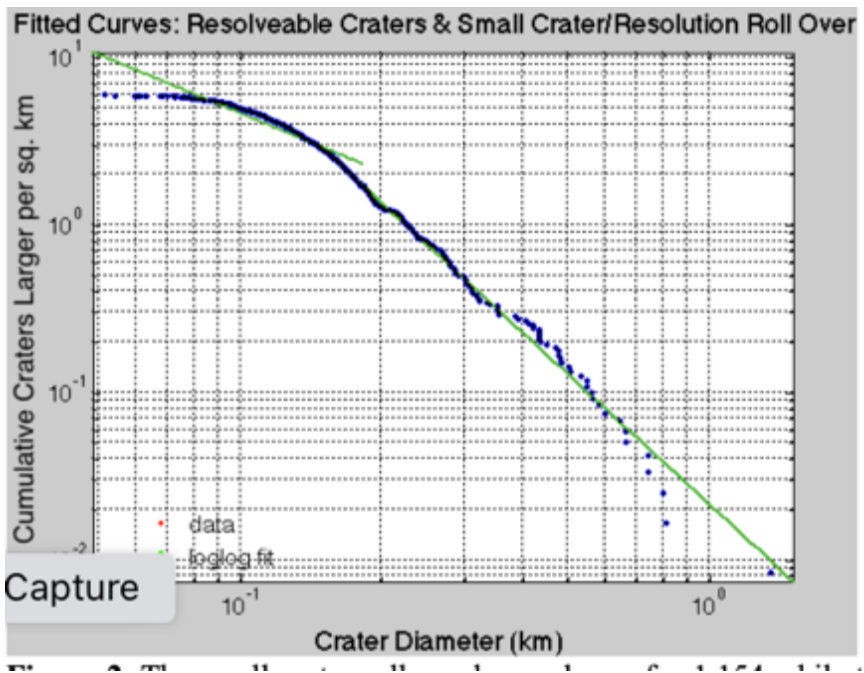

The simulation includes every crater (bins of one) while the observations have been binned more coarsely. The simulated impactor population is limited to objects larger than 4 meters, which accounts for the fall-off in craters at about 40 m in the counts. The data falls off below 100 m because of resolution. The simulation captures every crater over the entire asteroid surface, while the data was obtained from a 19% area patch of its surface. Therefore, the numbers from the observations were scaled up by a factor of 5.2. Of course, in that process for the data a few large craters could be overlooked.

The simulation produces about 12 craters larger than 1 km, with the largest being a few km diameter, but the largest size varies a lot between different simulations. The calculation does not include any effects of erasure or masking of small craters by the ejecta of large ones. However, since the Gaspra crater topology is relatively sparse, far less than cratering equilibrium, those effects would not be large.

The crater size distributions on a given asteroid surface are a rather direct mapping of a population SFD, but the ratio of those two is entirely determined by the crater scaling assumed. There are a lot of features about crater scaling not well determined. For example, in the present case it is unknown whether and how much effect there is of size-dependent strength. There are questions about whether and at what size spall cratering exists and if it may dominate over bowl shaped craters. However, the apparent 'bump" in the crater size distribution just below the 1 km size is very interesting, that is where the transition from bowl shaped to spall craters is predicted.

Spall cratering has not been considered in asteroid studies to date although its occurrence in terrestrial lab studies is inescapable. It may be important on small asteroids, those less than perhaps 20 km in diameter. A careful re-study of the surface cratering on such asteroids may furnish important information. Features such as a low number of small craters, and a large number of surface rocks (possible spall fragments) may be indicators of spall cratering. More study is required before it can be claimed with confidence that this spall cratering is in fact applicable, although the possibility is intriguing[28].

But a final warning. The result for a surface cratering age depends strongly on the assumed population of small impactors (which is very speculative) as well as on the crater scaling (which may be unknown to within a factor of ten). I have found that, within plausible variations, the estimated surface age of Gaspra can range from a few hundred Myr to a Byr. The uncertainty in the SFD and the crater scaling diminishes the accuracy of using surface crater density as a meaningful indicator of the surface cratering age for small asteroids.

---

[28] Bottke et al., 2020 wonder if scaling laws change depending on the size of craters. Indeed, that is the case for spall cratering for brittle targets and may be an important factor to include in their studies.

*SSAH* also produces movies of the cratered surface history. **Fig. 26** is a frame from the movie generated at the time of 190 Myr of all craters larger than 3 m. The crater locations were assigned randomly.

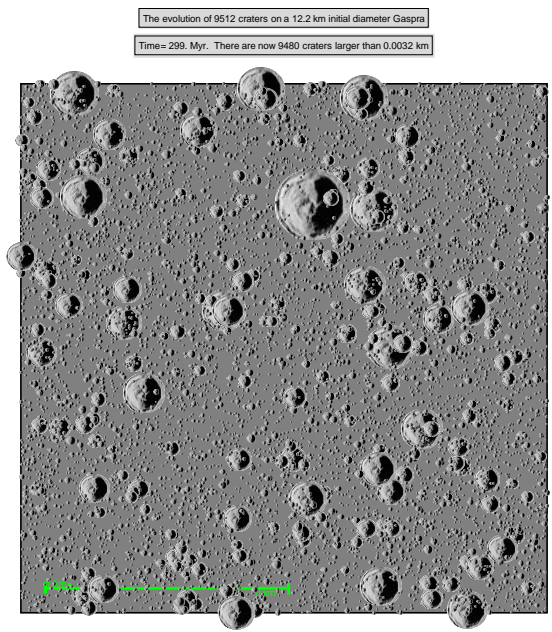

**Figure 26.** The cratered surface of Gaspra from a simulation at the time of 300 Myr, shown on a 22 x 22 km square. There are about 16 craters larger than 1 km, and the largest is about 2.5km, similar to some identified on Gaspra

## *7.3. Asteroid families*

The break-up and re-accumulation of a large asteroid results in a collection of dynamically and compositionally related fragments which are known as asteroid families. The younger ones, having had only limited evolution, provide data on the size distribution of those fragments. Understanding conditions that produce families is an important part of asteroid studies.

A simulation was run using *SSAH* to consider family formations. A set of target asteroids used 3380 objects with diameters greater than 100 km, ten times the current number of main belt objects larger than 100 km diameter. In the simulation, 186 objects or 5.5% were dispersed in 4 Byr. That would imply 18-19 families with a parent body >100 km. That number matches the estimate of 20 families[29] given in Bottke et al., 2005. That consistency also lends some support to the assumption of a steady population for several Byr.

There is another use for *SSAH* regarding families. While it cannot run histories in reverse time, one can make an educated guess about an initial size distribution and age, and then, via the simulation, determine what that distribution would be at the present time. An initial distribution can be found in that way by iterating a couple of times. This is the same process which was indicated above to adjust the diameters in a spin database so that the simulation will end, not start, with those diameters. One such study has been done for the Eos family, but results are not presented here. Of course, the biggest differences between initial and present family size distributions are for the smaller members.

---

[29] The difficulty in the observations is to determine the initial diameters of the parent bodies.

### 7.4. Remaining Lifetimes

Above in §2.8.4, estimates of the remaining lifetimes for various sizes were made by calculating the average time for which a catastrophic dispersion would occur. That has been a standard in the literature to date. However, the simulations here include additional detail and allow additional important features about lifetimes to be determined.

The first question that arises is the exact definition of an object's lifetime. The time until the first dispersion event defines an upper limit to the duration until the remaining mass is less than ½ of the original mass, based on assuming no previous mass loss. Other criteria can be used; for example Dohnanyi, 1969 used the time to ½ diameter (1/8 mass). However, since the remaining lifetime of an object already reduced to less than half mass is very short, the average lifetime based even upon a complete pulverization is only a little larger than that to ½ mass.

Further some objects, especially the smaller ones, may be eroded substantially, and occasionally even to less than half mass before a dispersion event. **Figs. 13** showed that fact. In his historic analysis Dohnanyi, 1969, using the scaling popular at that, time suggested that for a 1 km object 2% would be eroded to half mass before a CD. The number obtained in the simulations here is more like 10%, mostly from "near" CD events. But more important is the fact that their CD threshold is decreasing as they are shrinking in size. Here an object's lifetime is taken to be the duration before its mass is reduced to less than 1/2 of the original mass by any means (Henych and Holsapple, 2018).

To study how lifetimes depend on size, simulations of 1000 C-Type asteroids of a fixed size were made. The distribution of lifetimes for 1 km asteroids, based on the time for the reduction to less than ½ of its original mass, is presented in **Fig. 27**.

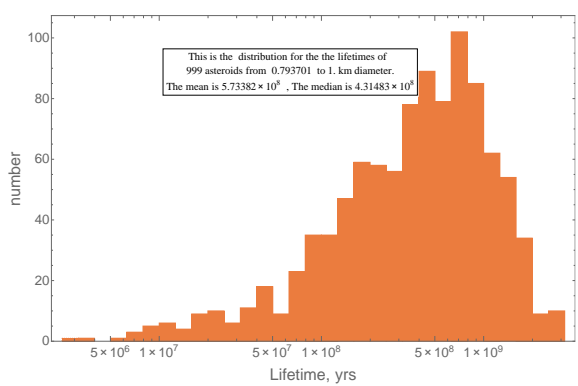

**Figure 27.** The distribution of the lifetimes to ½ mass of originally 1 km diameter C-Type asteroids. The lifetimes range essentially from a few 10 Myr to several Byr years. The mean lifetime is 0.57 Byr and the median is 0.43 Byr.

The distribution of ages is, as expected, approximately exponential since the events (large impacts) reducing their mass occur at a constant average rate. Half were reduced to less than half-mass by 430 Myr, and the average was at 570 Myr. Of the total of 1000 objects, 112 or 11% were reduce to less than half-mass by erosions alone.

These can be compared to the results of Table 1 which predicted an average lifetime of 1.4 Byr, based on the time to the first CD event. There are two reasons for this large difference,

First, in 11% of the cases, the reduction to half mass was from erosion alone, not CD. The second reason for that difference is apparent from **Fig. 28,** which show that 50% of the 1000 objects had a significant size reduction by erosion before a CD terminates their effective life. An 800 m asteroid has 60% more impactors in the population larger that the *d\*=33 m* impactor that will disperse it, which shortens its predicted life by a factor of 1.6.

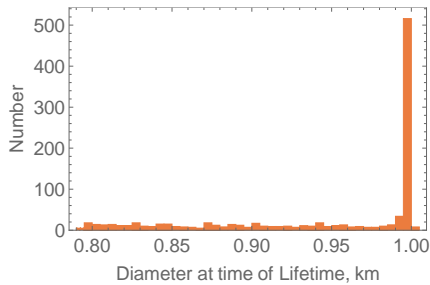

**Figure 28.** The diameters of 1000 1 km asteroid at the time they are diminished to less than half mass. 60% have substantial mass loss before catastrophic dispersions.

**Figure 29** shows similar results for lifetimes of other sizes of C-Type asteroids, compared to the simple estimates used in the literature and in Table 1 above. The actual lifetimes are factors of 2-3 lower for the reasons stated.

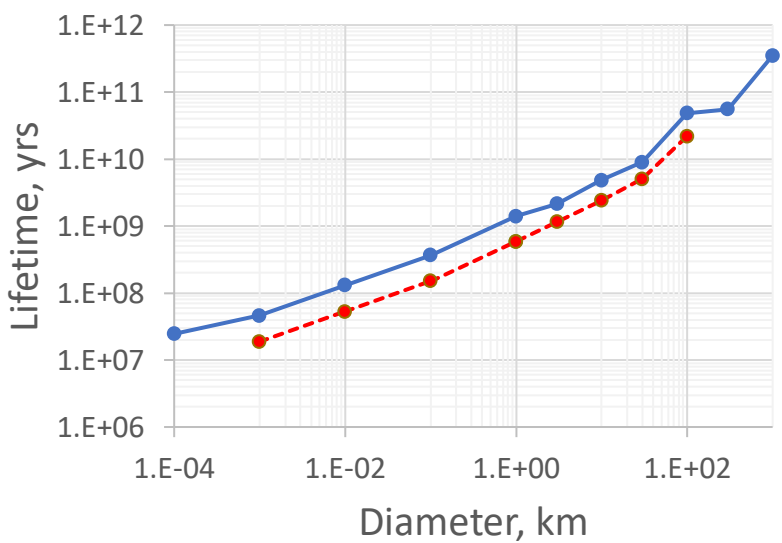

**Figure 29.** The lifetime estimates using only the time to a first dispersion using the original diameter (blue, solid curve), compared to a detailed simulation (red, dashed curve). The red curve accounts for a noticeable fraction reduced to half mass by erosion alone and a reduced diameter at the actual first dispersion event.

## 8. NEO VERSUS MAIN BELT DATA

The database being used here for the spin distribution comparison includes NEO's, with diameters mostly smaller than 1 km, and main belt bodies, with diameters mostly greater than 1 km. But there is significant overlap for sub km diameters. In the spin database, their spin magnitude distributions map continuously across the transition between them. The simulations of *SSAH* treat both sets of data as one, subjecting them to the entire simulation time of conditions in the main belt. But why should we expect the spin distributions of the NEOs to be the same as if they were exposed to the that time history in the main belt?

A possibility is straightforward. The smaller objects in the main belt are destroyed several times during the simulations. Those destroyed are replaced by new equal sized objects that are fragments from larger objects, and the spin distributions recycle; so that the spin distribution is essentially fixed in time after a short initialization phase, as was illustrated in **Figs. 19.** If, at some time, a small object moves from the main belt into the NEO space, the spin is likely fixed at that time. The NEOs are a random 'snapshot' of those objects in the main belt. Therefore, the distributions of spin would be the same, whether frozen at a fixed instant, or as a NEO. Of course, that ignores random close encounters to the inner planets which can change the spin.

However, perhaps the simulation using main belt conditions should not match the spins of the NEOs. A mismatch might be attributed to a possibility that the smallest fast spinners are not so easily diverted, so the NEOs are not a random sample of the main belt objects. Or perhaps conditions in the NEO space such as close encounters changed the spins. Or perhaps the data for the small fast spinners is incomplete, and an observational bias. I will leave those conjectures for others to judge. However, the apparent lack of differences for at least spin values and their distribution over the overlapping range of the two sets of data suggests that the first explanation is valid.

## 9. COLLISIONS AND YORP

The spin histories calculated in these simulations are from collision events alone. The simulations match very well with the data. That raises questions, because in the last two decades it has become popular to assert that YORP[30] effects are as, or perhaps even more, significant than collisions in the history of asteroids in the main belt. If the results here are correct, that opinion might be questioned.

It is recognized that the YORP is strongly dependent on the detailed state of an object, including its shape, surface roughness, spin, tilt, wobble, regolith albedo, mass distribution, and so forth. Questions about the broad acceptance of YORP effects have been given (Statler, 2009, 2015, Marzari et al., 2020). Each impact changes many of the above details and may entirely reset the YORP mechanism. However, as yet, we have no quantitative measures to connect specific changes to the so-called and largely unknown "YORP coefficients".

Three features in the data have been prominently mentioned as signatures of YORP: 1.) the observed excess of very fast and slow rotators in asteroids smaller than 40 km in size; 2.) spin-up effects including change of shape, binary formation, and asteroid pairs; 3) aligned tilts of principal axes in some families. But the first two of those occur in this study from impacts alone.

The terminology "excess of very fast and slow rotators" refers to comparisons of the distributions of spin to a Maxwellian distribution. The assumption that the spin distributions "should be" Maxwellian is inherent in many older spin studies. It is based on the view that the spin is a result of a very large number of random infinitesimal increments of spin, producing a random-walk process of slowly increasing spin. Therefore, data for asteroids such as the slowly rotating Mathilde, or for the rapidly rotating 2001 OE84 are interpreted as "outliers" and "puzzling". However, it has been known for some time that the spin of the asteroids is not primarily from many small impacts. Instead, they are mostly produced by a small number of large impacts. That is readily apparent in this study (e.g., see the **Figs. 14**). The resulting spin distributions are non-Maxwellian, with "an excess of slow and fast spins". Collisions alone produce the non-Maxwellian spin distributions, which obviates the need for an additional mechanism.

---

[30] Yarkovsky-O'Keefe-Radzievskii-Paddack (YORP) effect. See Rubincam, 2000.

Then there is the clustering of spin values for the range of 1 to 10 km objects just below the 10/day gravity spin limit, just the diameter range where YORP might apply. Spin-up over that limit is thought to be the source of top-shaped objects, and perhaps rotational fission into binaries and asteroid pairs. Indeed, that spin-up assumption is adopted here. But the results here contradict the statement that those processes require YORP. The results here produce those clusters using collisions alone[31]. Again, that feature does not necessarily imply a need for YORP.

So, while some of the inferences about YORP are contradicted here, that does not mean that YORP is not present. However, it may not be appropriate to assume that a YORP process can remain constant over any significant time. Instead, those processes may just add random fluctuations to the collisional spin histories calculated here.

It would be illuminating to add YORP processes to these simulations. However, it is very difficult to do that correctly. The problem is to correctly assess the change in the YORP coefficient at each impact; accounting for the detailed changes[32]. It may be the only viable way is to just assign YORP resets randomly.

## 10. SUMMARY AND CONCLUSIONS

A comprehensive study and a code '*SSAH*' to simulate the history of the asteroids in the main asteroid belt is presented. Their history is a result of many collisions, which in number and magnitude are determined by the environment of the main belt. The outcomes of the simulation of those impacts depend upon various models of the impact mechanics. I present and discuss models for erosion, cratering, catastrophic dispersions, spins, binaries, and tumblers.

*SSAH* calculates each individual history of a chosen set of sample targets. It uses stochastic measures for the parameters defining the sample asteroid, for the geometry of the impacts, for impactor properties and collisions. It accounts for the large number of impacts on a given target by using an explicit-implicit approach which calculates explicit detailed results for perhaps a thousand of the largest impactors and then grouped average results for the remainder. The impact models are defined in separate functions in *SSAH*, which allows easy modification for testing or if newer or better models are developed.

This manuscript focuses only on the single most important special case: the population of the main belt is assumed to be steady in time. I do not calculate the evolution of the population, which was the primary topic of the older asteroid history studies, although *SSAH* does have that capability. That evolution is determined by the balance between the models for fragmentation/mass reduction, and dynamic depletions, both of which are highly speculative. A steady population arises when fragments equally replace those bodies reduced in size or lost by impacts and by dynamic processes. In that case a simulation does not require time steps except to study the histories, because the histories of all of the objects are independent.

---

[31] Holsapple, 2010 presented an analysis of shape adjustments of a body that is slowly being spun up and used the word YORP in the title. However, in fact his analysis applies to spin-up by any mechanism. Now I regret using that title. Results here suggest the primary spin-up is from collisions, not YORP
[32] A recent approach in that direction is that of (Marzari et al., 2020). For impacts, they consider spin-up, using the 'sticky impactor' assumption, and do not consider further details. They do not calculate mass depletion. After each large impact, they reassign the YORP coefficients in a random way.

Many informative results are found, including the distributions of the asteroid spins, the divisions between gravity-limited spins and the small fast-spinners, the role of strength for the fast spinners, possible processes of the formation of binaries, the formation and subsequent relaxation of tumbling objects, the lifetimes of the asteroids, the formation and evolution of families, and the crater distributions on asteroids. Many connections to past studies and significant new insights are presented.

In summary:

1). The general nature of the observed spin distributions is fully simulated.

2). The size and spins of the largest diameter asteroids are primordial. There are not enough large impactors to affect their spins over the life of the solar system.

3) The observed "V-shape" minimum in the spin distribution is the intersection of a primordial spin distribution with the distribution caused by 4.5 Byr of impacts.

4). The erosion and spins of the asteroids are not the accumulation of a multitude of infinitesimal changes. Instead, they are caused by a few large events. Models based on the 'small increment, collisionally equilibrated' assumption are not valid

5). YORP effects are not needed to explain the spin histories.

6). Asteroids are spun up to the gravity limit and may create binaries by collisions alone, YORP is not required.

7). Asteroid rotation can be slowed or even be reversed in a single impact. A few slow spinners such as Mathilde can be expected. However, on average, impacts increase spin.

8). The spins of objects in the transition region between 200 m and 2 km is a transition between gravity-dominated and strength-dominated impacts. A reforming of a configuration may change the response to a subsequent limit spin. The nature of those readjustments and the statistics of the distribution of spins in that area requires further study.

9). The distribution of the small fast spinners is not affected in any way by measures of the material strength of those objects. That data does not give any meaningful information about the strength of those objects. Their states are a simple result of the balance between spin-up size reduction for impacts into the small objects.

10). The spins for all except the very largest objects are in equilibrium between effects of spin-up and catastrophic dispersions. For that reason, the resulting spin distributions are largely unaffected by the population of impactors.

11). For a larger population, each object will have spin changes at a faster rate, but also will be dispersed at a faster rate. Therefore, it is not possible to solve for over-all population magnitudes using the present spin data.

12). Wobbling and tumbling states are much more complicated than a simple decay of an initially tumbling asteroid to some small angle. A complete theory for the history of wobbling is derived. The decay time depends on the initial wobble and on the internal dissipation, which is not well defined. During that time, many more changes in wobbling will occur from subsequent impacts.

13). Cratering on small S-Type asteroids may be dominated by the spall cratering process which has not been considered to date. Further, the crater surface lifetime found depends so much on the unknown populations of small impactors and on the crater scaling that it is quite unreliable.

14). Lifetimes calculated using the previous literature measure as the time to a first expected catastrophic dispersion are too large by factors of several. Actual lifetimes are strongly affected by erosion; both erosions that reduce an object to less than 1/2 of its original mass and prior erosion effects that reduce the difficulty of catastrophic dispersion.

15). There are a number of different areas using models that are not totally determined, to different degrees. Plausible guesses can be made as starting points, and the code allows one to study the effects of changing the models on the simulations. This study helps identify those. Both the models and the data will improve in the future.

The code *SSAH* is available in the public domain [Here](https://www.dropbox.com/sh/3owgbo376q50fpp/AACmeJatGL7pZjL1BU7bATbra?dl=0) (https://www.dropbox.com/sh/3owgbo376q50fpp/AACmeJatGL7pZjL1BU7bATbra?dl=0). One should download the entire contents of the folder 'SSAH Folder'. I encourage others to use *SSAH* to further investigate the asteroid belt history. Please give credit to this author on studies making extensive use of *SSAH*.

## 11. REFERENCES.

## 12. ACKNOWLEDGEMENTS

# APPENDIX A. THE *SSAH* CODE STRUCTURE

*SSAH* ("Stochastic Simulations of Asteroid Histories") is a code that constructs the history of asteroids in the main asteroid belt using a Monte-Carlo stochastic approach. It is written in the 'Wolfram code' of *Mathematica* and requires the *Mathematica* program to run. It is designed to be a general tool for research and analysis of asteroids in the main belt.

The goal was to write a comprehensive code using current physical models for the impact processes in the main belt. It is designed with a simple structure and outline. All of the physical models are included in separate functions so a user can easily revise them as new ones arrive; and can easily making changes and/or improvements to those models to observe their effect on the outcomes, and to compare to known data.

## A.1. Inputs

The main input choices are made in graphical windows, as shown below in **Fig. A.1**:

**Figure A.1.** The panels for input choices. Note the buttons that present plots of the significant models used.

Each input has an adjacent button labeled *'Huh?'* which explains the meaning of that input.

The code is segmented into 8 parts ('cells' in *Mathematica* terminology) which automatically run sequentially. The first six cells define the inputs, define all of the functions, open input files and initialize variables. The primary calculation loops are in the 7[th] cell. Those are defined in detail in the Appendices O and L. Finally, the last cell constructs and displays up to 13 specific plots, some with post-output user-adjustable features.

### A.2. Time Stepping, Targets and Impactors

A set of sample targets is chosen. Each target has a diameter and is assigned a composition according to input choices. Presently the choices include all S-type, all C-type, or a 25/75% combination. Each type has predefined properties of strength versus size, impact scaling exponent $\mu$, density, CD curve, and the ejecta and cratering coefficients as required for the models.

A total simulation time and incremental time steps are defined. Over each time step the impactor population is assumed to be constant. If a non-steady population is desired, then the sets of fragments and dynamic depletions over the time step are saved, and they are used to update the population at the end of each time step. If a steady population is chosen, then those time steps are only used to make plots of results at each step for later movies. If the "reborn" flag is set, then when a target is destroyed, a new one with the same initial diameter and random spin is inserted to take its place. That essentially assumes that for a steady population, each object lost is replenished by a fragment from a larger impacted object.

In each time step, and for each target, there is a calculation loop on targets. A binned list of impactors to be encountered is constructed. That list is chosen from a binomial distribution, where the expected number of impactors in a time step $\Delta t$ is simply the total population in each bin times a ratio of the volume swept out by the target in that time step compared to the total assumed volume of the entire main belt. The swept volume is $(\pi d^2)/4) U \Delta t$ and the total belt volume $W$ is determined from the *intrinsic collision probability* $P_i$ (Kuiper, 1950) as $W = \frac{\pi U}{P_i}$, so that the volume ratio is $P_i \left(\frac{d^2}{4}\right) \Delta t$ (see §2.4). The expected number of impactors in each bin is rounded to the nearest integer. According to the law of large numbers, when the number of impactors in a bin is large, the number of impactors in that bin is just the volume ratio times the total population number in that bin.

### A.2.1. The Impactor Loops: Explicit and Implicit Calculations

For each target the impactor binned list ranges in diameters from a few largest ones to a very large number of smaller ones in higher numbered bins. A loop over those impactors is defined to account for the impacts of those impactors. One cannot explicitly calculate each impact for millions of impactors. Thus, the outcomes for impacts is constructed in two parts. That is the key construct here making the simulation feasible.

An impactor with an energy greater than 10[-3] of the energy corresponding to target's CD threshold $Q^*$ (using an average velocity and angle) is deemed 'significant'. Then each such impactor is assigned random parameters: a specific random diameter according to the bin it is in, a random explicit impact time within the defined time step, a random velocity according to the distribution shown in Sec. 2.3 above, a type, and angles $\{\phi, \Phi, \theta, \Theta\}$ defining the impact geometry, as defined in §3.2. Typically, each target has about 1000 explicit impacts.

Then the mass loss and spin increment are calculated and recorded in turn for each explicit impact. The mass loss and resulting diminished diameter are scalar changes; while each increment of spin has three components defined in the target's principal axes system. If a substantial component of the new spin is not about the principal axis, the target is flagged as a "wobbler". The amount of that wobble is tracked through all subsequent impacts and relaxation periods, as described in §4, and is identified on outputs. The angle misalignment to recognize a wobbler is chosen in a user 'knob', the default is 15˚, with post-calculation changes possible in the plots.

The net effect of impacts from all of the remaining small impactors is determined by an implicit calculation. For each bin, the average effect using an average impactor diameter, geometry parameters and velocity is multiplied by the number of implicit impactors in that bin. This two-part explicit-implicit method is very robust and efficient.

### *A.3. Outputs*

In *SSAH* there are 13 plots of results. The first 4 show the resultant final states of the surviving objects, including the entire spin-size distribution, the running mean of that data compared to that of the database, a histogram of the spins in a user-selected diameter range, and a histogram of numbers reduced to less than half mass by either erosion or CD (lifetimes).

The next three show the time histories of the diameter, mass, spin, and wobble of a pre-chosen number of asteroids. Then there are plots of final wobble values, time-slice movies of the spin-diameter paths, the entire spin-diameter path, and a slide-show in time of the spin-diameter distribution and mean spins. In addition, when there is a single target, plots of the crater count population and a simulated movie of the cratered surface are provided.